\documentclass[aps, prb,twocolumn,groupedaddress,showpacs]{revtex4-1}

\usepackage{graphicx,xcolor}
\usepackage{hyperref}
\usepackage{amssymb}
\usepackage{amsmath}
\usepackage{mathrsfs}
\hypersetup{
    colorlinks,
    linkcolor={blue!},
    citecolor={blue!},
    urlcolor={blue!}
}
\usepackage{braket}

\begin{document}

\title{Andreev bound states probed in three-terminal quantum dots}

\author{J. Gramich}
\author{A. Baumgartner}
\email{andreas.baumgartner@unibas.ch}
\author{C. Sch\"onenberger}
\affiliation{Department of Physics, University of Basel, Klingelbergstrasse 82, CH-4056 Basel, Switzerland}

\date{\today}

\begin{abstract}
Andreev bound states (ABSs) are well-defined many-body quantum states that emerge from the hybridization of individual quantum dot (QD) states with a superconductor and exhibit very rich and fundamental phenomena. We demonstrate several new electron transport phenomena mediated by ABSs that form on three-terminal carbon nanotube (CNT) QDs, with one superconducting (S) contact in the center and two adjacent normal metal (N) contacts. Three-terminal spectroscopy allows us to identify the coupling to the N contacts as the origin of the Andreev resonance (AR) linewidths and to determine the critical coupling strengths to S, for which a ground state (or quantum phase) transition in such S-QD systems can occur. In addition, we ascribe replicas of the lowest-energy ABS resonance to transitions between the ABS and odd-parity excited QD states, a process we call excited state ABS resonances. In the conductance between the two N contacts we find a characteristic pattern of positive and negative differential subgap conductance, which we explain by considering two nonlocal processes, the creation of Cooper pairs in S by electrons from both N terminals, and a novel transport mechanism called resonant ABS tunneling, possible only in multi-terminal QD devices. In the latter process, electrons are transferred via the ABS without effectively creating Cooper pairs in S. The three-terminal geometry also allows spectroscopy experiments with different boundary conditions, for example by leaving S floating. Surprisingly, we find that, depending on the boundary conditions and the device parameters, the experiments either show single-particle Coulomb blockade resonances, ABS characteristics, or both in the same measurements, seemingly contradicting the notion of ABSs replacing the single particle states as eigenstates of the QD. We qualitatively explain these results as originating from the finite time scale required for the coherent oscillations between the superposition states after a single electron tunneling event. These experiments demonstrate that three-terminal experiments on a single complex quantum object can also be useful to investigate charge dynamics otherwise not accessible due to the very high frequencies.
\end{abstract}


\maketitle

\section{\label{sec:Intro}Introduction}

Nanoscale electronic devices in contact with superconducting contacts (S) exhibit a large variety of fundamental physical phenomena and play, for example, a central role in schemes for quantum computation.\cite{DeFranceschi:2010,Sarma:2015} If S is in contact with the many quantum channels of a normal metal (N), phase-coherent Andreev reflections lead to electron pairing and an induced superconducting gap in N. This proximity effect has recently been demonstrated also for one-dimensional semiconducting nanowires,\cite{Chang:2015} where strong spin-orbit interactions can give rise to Majorana bound states.\cite{Mourik:2012,Stanescu:2013a,Albrecht:2016} If S is strongly coupled to a single channel quantum dot (QD), new subgap eigenstates form, which are known as Andreev bound states (ABS).\cite{Eichler:2007,Deacon:2010,Pillet:2010,Dirks:2011} ABSs carry the supercurrent in Josephson junctions,\cite{Doh:2005,vanDam:2006,Jarillo-Herrero:2006,Jorgensen:2007,Eichler:2009,Maurand:2012,Delagrange:2015,Delagrange:2016} and thus constitute a model system to investigate the superconducting proximity effect in QDs.\cite{Martin-Rodero:2011} ABSs might also be exploited as Andreev quantum bits \cite{Zazunov:2003,Janvier:2015} and have recently attracted considerable attention in both, theoretical \cite{Vecino:2003,Bauer:2007,Meng:2009,Futterer:2009,Eldridge:2010,Braggio:2011,Martin-Rodero:2011,Michalek:2015,Trocha:2015,Zitko:2015} and experimental work.\cite{Deacon:2010,Pillet:2010,Dirks:2011,Bretheau:2013,Kim:2013,Pillet:2013,Kumar:2014,Lee:2014,Schindele:2014,Higginbotham:2015,Janvier:2015}

Previous experiments on ABSs were focused on QDs with two contacts,\cite{Deacon:2010,Pillet:2010,Dirks:2011,Bretheau:2013,Kim:2013,Pillet:2013,Kumar:2014,Lee:2014} and only few were possible in multi-terminal devices.\cite{Schindele:2014,Higginbotham:2015} Three-terminal ABS devices allow for new transport mechanisms, so that nonlocal processes like Cooper pair splitting\cite{Hofstetter:2009, Herrmann_Kontos_Strunk_PRL104_2010, Schindele:2012, Fulop_Baumgartner_PRL115_2015} compete with local mechanisms like Andreev tunneling.\cite{GramichPRL:2015} Such mechanisms are expected to result in new effects like the triplet blockade \cite{Eldridge:2010}, characteristic patterns in the electrical conductance,\cite{Futterer:2009,Michalek:2015} and other nonlocal effects.\cite{Trocha:2015,Zitko:2015} In addition, even the most basic ABS characteristics are expected to be determined by the barrier strengths, such as the ABS resonance broadening\cite{Pillet:2010,Lee:2014} and whether a quantum phase transition in the S-QD many-body ground state occurs as a function of the gate and bias voltages. However, the coupling strengths can only be accessed unambiguously in three terminal geometries.\cite{Leturcq:2004,Leturcq:2005,Jacobsen:2012,Jacobsen:2014}

\begin{figure}[b]
	\begin{center}
	\includegraphics[width=\columnwidth]{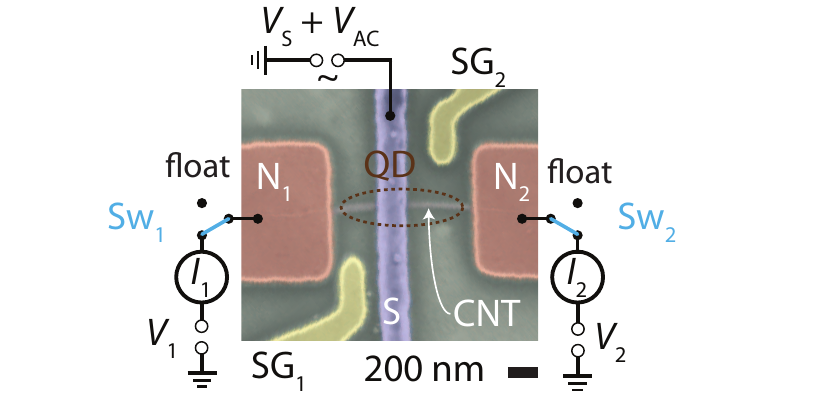}
    \end{center} 	
	\caption{(Color online) False-color SEM image of a typical device and schematic of the measurement setup.}
	\label{Fig1}
\end{figure}

Here we report experiments on three-terminal QD devices fabricated on carbon nanotubes (CNTs) with one central superconducting contact inducing ABSs on the QD. A scanning electron microscopy (SEM) image is shown in Fig.~\ref{Fig1}. We investigate three devices with different coupling strengths $\Gamma_{\rm S}$ to S to investigate several of these open questions and demonstrate a series of new effects and measurement configurations. We use Pb as superconductor, which results in a large energy gap $\Delta$ and thus in a high relative spectroscopic resolution,\cite{GramichAPL:2016} crucial for our experiments. The paper is structured as follows: In Sec.~\ref{sec:ABS_basics} we summarize our intuitive picture of ABSs and ABS-mediated electron transport in two-terminal devices. Sec.~\ref{sec:Fab} describes the sample fabrication and measurement setup. Then we characterize each of the investigated devices by transport spectroscopy in Sec.~\ref{sec:ABS-devicecharacteriation}, from which we extract the tunnel coupling strengths to the individual contacts and demonstrate that $\Gamma_{\rm S}$ determines the system ground state and ABS dispersion, while the coupling strengths to the normal metal contacts, $\Gamma_\mathrm{1,2}$, determine the spectroscopic width of the Andreev resonances (ARs).\cite{Martin-Rodero:2011,Michalek:2015} One device shows replicas of the lowest-energy ABS resonance at higher energies, which we ascribe to transitions to excited odd-parity QD states in Sec.~\ref{sec:ABS-Addbias-ExcitedAR}. In Sec.~\ref{sec:ABS-SerialTransport}, we analyze the transport between the two normal metal contacts mediated by ABSs, in which competing local and nonlocal transport mechanisms give rise to a characteristic pattern of positive and negative differential conductance. These findings are well captured by a simple rate equation model, which allows us to identify Cooper pair splitting\cite{Hofstetter:2009, Schindele:2012} and a new three-terminal subgap process we call {\it resonant ABS tunneling}. In Sec.~\ref{sec:ABS-FloatingS-Proximity} we show that depending on the boundary conditions imposed in the experiments, the measured conductance either exhibits ABS or Coulomb blockade characteristics, or both at the same time, which we tentatively attribute to finite frequency coherent oscillations between the single particle basis states when an ABS is excited by single electron tunneling.

\section{\label{sec:ABS_basics}ABS mediated two-terminal transport}

For illustration and to establish the terms used below, we consider a QD with a single, spin-degenerate level strongly tunnel coupled to a superconductor with a strength $\Gamma_{\rm S}$, and weakly to a normal contact. The former can couple even charge states of the isolated QD by exchanging Cooper pairs to form new eigenstates of the S-QD system. For $\Delta \rightarrow \infty$ (superconducting atomic limit), the emerging ABS can be written as a superposition of the even (empty and doubly occupied) charge states of the isolated QD, i.e., as a singlet eigenstate $\ket{-}=u\ket{0}-v^{*}\ket{\uparrow \downarrow}$, with $u$ and $v$ the gate-dependent Bogoliubov- de~Gennes (BdG) amplitudes and $E_{-}$ the corresponding eigenenergy. The orthogonal eigenstate at higher energies reads $\ket{+}=v\ket{0}+u^{*}\ket{\uparrow \downarrow}$. In contrast, the odd charge states remain unperturbed doublets $\ket{\sigma}$, with $\sigma=\{\uparrow,\downarrow\}$ and energies $E_\sigma$.\cite{Vecino:2003,Bauer:2007,Meng:2009,Braggio:2011,Martin-Rodero:2011} 
We note that a similarly intuitive picture can be drawn in the limit of small $\Delta$, in which unpaired QD electrons can form Yu-Shiba-Rusinov singlet states with quasi-particles in the superconductor at energies below the gap and with transport processes formally very similar to the ones discussed below.\cite{Chang_Nygard_Marcus_PRL110_2013, Jellinggaard_Grove-Rasmussen_Nygard_PRB94_2016} Here we choose the large-$\Delta$ limit for the discussion because of its simple analytical expressions.

\begin{figure}[t]
	\begin{center}
	\includegraphics[width=\columnwidth]{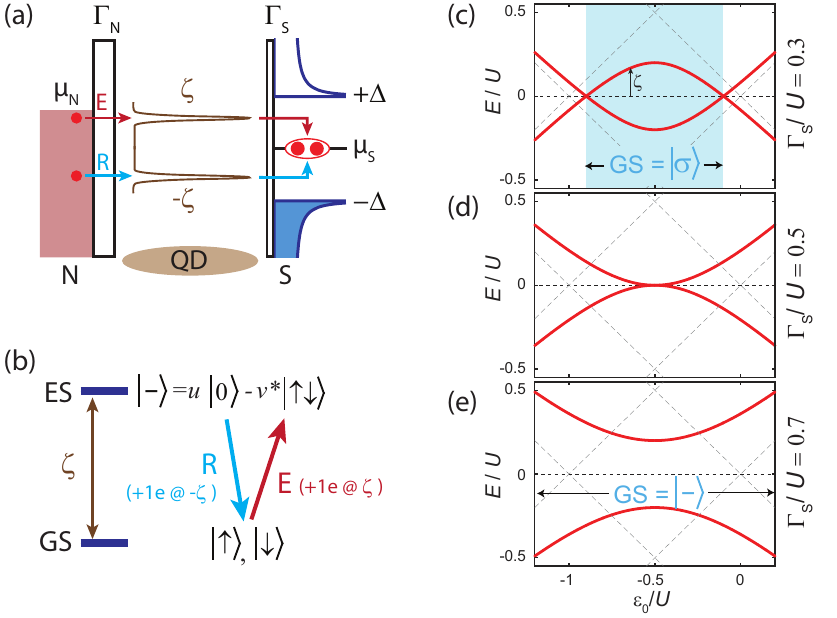}
    \end{center} 	
	\caption{(Color online) (a) Schematic of an ABS-mediated two-terminal transport process: an electron at the energy $\zeta$ excites the S-QD system (E) and a subsequent electron at $-\zeta$ allows the system to relax again (R), thereby adding a Cooper pair to S. (b) Energy diagram of the excitation and relaxation processes. (c)-(e) ABS resonance energy $\zeta$ for $\Delta\rightarrow\infty$ as a function of the level position $\epsilon_0$ for $\Gamma_{\rm S}/U=0.3$, $0.5$ and $0.7$, respectively.}
	\label{Fig2}
\end{figure}

The low-energy excitation spectrum of the S-QD system in the large-$\Delta$ limit is depicted schematically in Figs.~\ref{Fig2}(a) and (b), for the case of the doublet being the ground state (GS), and the ABS the excited state (ES). Due to the different parity of the doublet and the ABS, one can directly probe the excitation spectrum by single electron spectroscopy using a weakly coupled normal metal tunnel probe (N) in an N-QD-S geometry.\cite{Deacon:2010,Dirks:2011,Lee:2014,Schindele:2014} A current through such a device is only possible if the bias matches the Andreev addition energy $\zeta=|E_{-}-E_\sigma|$, i.e. if the electrochemical potential $\mu_\mathrm{N}$ of N exceeds (lies below) the excitation energy $+\zeta$ ($-\zeta$) with respect to the electrochemical potential in S, $\mu_{\rm S}$. As illustrated in Figs.~\ref{Fig2}(a) and (b), charge is transported by an electron tunneling from N into the $\ket{\uparrow \downarrow}$ part of $\ket{-}$, which excites (E) the S-QD system from the GS $\ket{\sigma}$ to the ES $\ket{-}$. In a two-terminal device, the system can relax (R) back to the GS only by absorbing a second electron from N at a negative energy $-\zeta$ tunneling into the $\ket{0}$ part of $\ket{-}$, and the transfer of a Cooper pair to S. The probability of this process cycle scales with $u^2v^2$.\cite{Schindele:2014} For a reversed bias, a Cooper pair is removed from S. Since in this process two electrons are transferred from N to S we call the resulting subgap conductance features \textit{Andreev resonances} (ARs), which can be seen as a generalization of Andreev reflection found in non-interacting systems.

The ARs show a characteristic gate and bias dependence in the shape of "loops" that directly reflect the competition between superconducting pairing and the Coulomb repulsion on the QD.\cite{Martin-Rodero:2011} While deep in the even QD charge states the GS of the system is the singlet $\ket{-}$ ($E_- < E_\sigma$), in the odd charge state the charging energy $U$ favors the doublet $\ket{\sigma}$ as GS  ($E_\sigma < E_-$), in competition with the pairing induced by $\Gamma_\mathrm{S}$. As illustrated for the limit $\Delta\rightarrow\infty$ in Figs.~\ref{Fig2}(c)-(e), one finds for small enough $\Gamma_{\rm S}$ [Fig.~\ref{Fig2}(c)] a gate-tunable transition from $\ket{-}$ to $\ket{\sigma}$ at gate voltages where the AR loops intersect ($\zeta=0$). In contrast, for large $\Gamma_{\rm S}$ the AR loops do not intersect and the GS is always a singlet, as shown in Fig.~\ref{Fig2}(e). In the large-gap limit $\Delta/\Gamma_{\rm S}\rightarrow\infty$ the critical parameter is $\Gamma_\mathrm{S}/U$ with a GS transition occurring for $\Gamma_\mathrm{S}/U < 0.5$,\cite{Vecino:2003,Tanaka:2007,Bauer:2007,Meng:2009,Martin-Rodero:2011,Baranski:2013} see Fig.~\ref{Fig2}(d). The limit for a phase transition should be reduced for finite $\Delta$ due to a competition between the superconducting pairing and the formation of a Kondo singlet with the quasi-particles at $\Delta$.\cite{Bauer:2007,Meng:2009} In the strong-coupling limit with $\Delta/\Gamma_{\rm S}\lesssim 0.5$ theoretical results suggest the following relation for a GS transition to occur:
\begin{equation}
k_{\rm B}T_\mathrm{KS}/\Delta \lesssim C
\label{Kondo_Temp}
\end{equation}
where the Kondo temperature $k_{\rm B}T_\mathrm{KS}\approx\sqrt{\frac{U \Gamma_{\rm S}}{2}}\exp\left(-\frac{\pi U}{8\Gamma_{\rm S}} \right)$ characterizes the coupling to S at the electron-hole symmetry point. The number $C$ varies considerably in the literature, between $C\approx 0.4$ and $C\approx 0.79$ in Ref.~\citenum{Sellier_PRB_2005}, $C\approx 0.6$ in Ref.~\citenum{Lee_deFranceschi_arxiv} and $C\approx 0.73$ in Ref.~\citenum{Bauer:2007} (values corrected for different definitions of $T_{\rm K}$). In the normal state we expect that the Kondo temperature, $T_{\rm K}$, is obtained by replacing $\Gamma_{\rm S}$ by the total coupling $\Gamma=\Gamma_1+\Gamma_2+\Gamma_{\rm S}$, where we already included the coupling to a third normal metal terminal. Though not experimentally demonstrated, yet, our intuitive picture suggests that the spectroscopic width of an AR line in the superconducting state at a bias $eV\ll\Delta$  is determined by the life-time of the excited state and thus by the finite coupling strength of the QD to all normal terminals, $\Gamma_\mathrm{N}=\Gamma_1+\Gamma_2$, see also Refs.~\citenum{Pillet:2010, Michalek:2015}.

\section{\label{sec:Fab}Device fabrication and measurment setup}


We have fabricated CNT three-terminal devices with a lead (Pb) based central $\sim 200\,$nm wide Pd/Pb/In ($4.5-6/110/20\,$nm) S contact,\cite{GramichAPL:2016} two Pd N contacts and two sidegates (SGs) using  close to residue free electron beam lithography.\cite{Samm_JAP_2014} The typical critical perpendicular field for these S contacts is $\sim 200\,$mT and we apply $400\,$mT for our normal state experiments. The highly p-doped Si/SiO$_2$ substrate serves as a backgate (BG). An SEM image of a typical device is shown in Fig.~\ref{Fig1}, which also illustrates our measurement setup: a dc voltage, $V_\mathrm{S}$, with a superimposed ac modulation of $\delta V_\mathrm{AC}=10\,\rm \mu V$ is applied to S, while measuring the variations $\delta I_{1,2}$ in the currents $I_{1,2}$ through the normal terminals N1 and N2 using standard lock-in techniques, which results in the differential conductances $G_\mathrm{1,2}=\delta I_{1,2}/\delta V_{\rm AC}$. In addition, we can set the potentials on N1 and N2 by the voltages $V_{1}$ and $V_{2}$, respectively (not shown), or leave them floating individually by the switches Sw$_{1(2)}$ outside the cryostat. Similarly, we can also apply the bias to N1 while measuring the current variations in S and N2, or leaving S floating. All measurements were performed in a dilution refrigerator at a base temperature of $\sim 35\,$mK.

We discuss three CNT devices A, B and C, each exhibiting subgap ARs. In device A, the Pd wetting layer was $4.5\,$nm thick, while in the devices B and C it was $6\,$nm. The room temperature resistances between an N terminal and S were $\sim 30-40\,\rm k\Omega $ (sample A) and $\sim 20\,\rm k\Omega$ (samples B and C).

\section{\label{sec:ABS-devicecharacteriation} S-QD ground state and life times}

In this section we investigate ABS mediated transport between N and S terminals in the three devices and take full advantage of the three-terminal geometry. Primarily, we focus on extracting the superconducting energy gap $\Delta$, the charging energy $U$ and the individual tunnel coupling strengths of the three contacts to the QD and and on relating them to the observed AR resonance patterns and the AR broadening. The relevant data for device A are shown in Fig.~\ref{fig:ABS-A-characterization}, device B in Fig.~\ref{fig:ABS-B-characterization} and device C in Fig.~\ref{fig:ABS-C-characterization} and will be discussed in the following subsections. In all figures we plot the differential conductances $G_1$ (left, a and c) and $G_2$ (right, b and d) as a function of a gate voltage and the bias applied to S, $V_\mathrm{S}$, while keeping $V_1=V_2 = 0$ and $V_{\rm BG}=0$. The top panels (a and b) show the data for S in the normal state obtained by applying a magnetic field of $B=0.4\,$T perpendicular to the substrate, and the bottom panels (c and d) the measurements with S in the superconducting state at $B=0$. The relevant extracted and derived device parameters are summarized in Table~\ref{tab:ABS-devices}.

\subsection{Device A}

\begin{figure}[t]
	\begin{center}
	\includegraphics[width=\columnwidth]{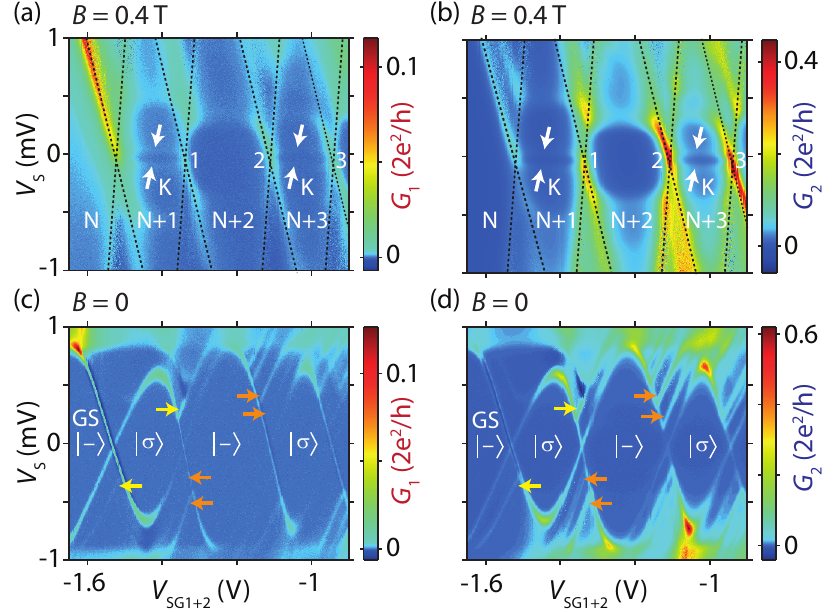}
    \end{center} 	
	\caption{(Color online) (a-d) Differential conductances $G_1$ and $G_2$ of device A as a function of the bias applied to S, $V_\mathrm{S}$, and the voltage $V_\mathrm{SG1+2}$ applied to both sidegates, for $V_\mathrm{BG}=0$ and the external magnetic field $B=0.4\,$T. (a-b)  or $B=0$ (c-d). The white arrows in (a-b) point out Zeeman-split Kondo resonances (K) and the numbers indicate the zero-bias CB resonances used for fitting. Yellow and orange arrows highlight excited state Andreev resonances.}
	\label{fig:ABS-A-characterization}
\end{figure}

In App.~\ref{app:SupplABS} we demonstrate in detail that the transport between all three terminals of device A is governed by a single QD. This is also evident in Figs.~\ref{fig:ABS-A-characterization}(a-b), which show the same Coulomb blockade (CB) diamond pattern (dashed lines) in $G_1$ and $G_2$ as a function of the bias and the voltage $V_\mathrm{SG1+2}=V_\mathrm{SG1}=V_\mathrm{SG2}$ applied to both SGs, with an even-odd shell-filling sequence in both simultaneously recorded conductance maps. In the odd diamonds we find horizontal conductance ridges (K, white arrows) due to Zeeman-split Kondo resonances.\cite{Nygard:2000,Babic:2004b} For increasing magnetic fields (not shown), we observe a Zeeman-splitting of these ridges with a g-factor of $g\approx 2.1$. Using additional measurement configurations,\cite{Leturcq:2005} e.g. with the bias applied to an N terminal, we find that the Kondo resonance solely originates from the S contact (not shown)

We first analyze the normal state data in Figs.~\ref{fig:ABS-A-characterization}(a) and (b). From the CB resonances we deduce a leverarm $\alpha_\mathrm{SG1+2}\approx 0.01$ of the combined sidegates, a charging energy of $U\approx 2.5\,$meV, and from excited state and inelastic cotunneling lines the lowest two orbital energies $\sim 0.3\,$meV and $\sim 0.5\,$meV. The three-terminal geometry allows one to determine the coupling strengths $\Gamma_i$ of the individual contacts to the QD by fitting a Breit-Wigner lineshape\cite{Sanchez:2005} to different two-terminal conductances.\cite{Leturcq:2004} This procedure is explained in more detail in App.~\ref{app:SupplABS}. From these fits, we obtain $\Gamma = \Gamma_1+\Gamma_2+\Gamma_\mathrm{S} = 204\,\mathrm{\mu eV}$, $\Gamma_\mathrm{S}=150\,\mathrm{\mu eV}$, $\Gamma_{1}=1\,\mathrm{\mu eV}$ and $\Gamma_{2}=53\,\mathrm{\mu eV}$ and thus $\Gamma_{\rm N}=\Gamma_1+\Gamma_2=54\,\mu$eV for resonance 1 in Fig.~\ref{fig:ABS-A-characterization}(a). For resonance 2 we obtain $\Gamma = 220\,\mathrm{\mu eV}$, $\Gamma_\mathrm{S}=173\,\mathrm{\mu eV}$, $\Gamma_{1}=1\,\mathrm{\mu eV}$ and $\Gamma_{2}=46\,\mathrm{\mu eV}$ and thus $\Gamma_{\rm N}=47\,\mu$eV, and for resonance 3 $\Gamma = 151\,\mathrm{\mu eV}$, $\Gamma_\mathrm{S}=118\,\mathrm{\mu eV}$, $\Gamma_{1}=3\,\mathrm{\mu eV}$ and $\Gamma_{2}=30\,\mathrm{\mu eV}$, so that $\Gamma_{\rm N}=33\,\mu$eV. We note that generally $\Gamma_\mathrm{S}\gg \Gamma_\mathrm{N}$. We estimate the Kondo temperature using $k_{\rm B}T_\mathrm{K}\approx\sqrt{\frac{U \Gamma}{2}}\exp\left(-\frac{\pi U}{8\Gamma} \right)$ at the electron-hole symmetry point in the middle of the odd diamonds,\cite{Haldane:1978,Deacon:2010,Kim:2013} which, for example, results for resonance 2 in $k_{\rm B}T_\mathrm{K} \approx 6\,\mathrm{\mu eV}$ ($T_\mathrm{K}\approx 70\,$mK), expected to be dominated by the temperature broadening.\cite{Zhang_Kern_NatureComm_2013}

If S is superconducting, as shown in Figs.~\ref{fig:ABS-A-characterization}(c-d), $G_1$ and $G_2$ both exhibit a gap $\Delta\approx 0.95\,$meV, with pronounced subgap resonances for $|eV_{\rm S}|\leq\Delta$. Also here we find the same conductance features (though of differing amplitudes) in $G_1$ and $G_2$. We discuss the higher energy replicas of the low energy AR pointed out by yellow and orange arrows in Figs.~\ref{fig:ABS-A-characterization}(c-d) in Sec.~\ref{sec:ABS-Addbias-ExcitedAR}. The subgap features are fully consistent with ARs: for even spin states (no Kondo ridge in the normal state) we observe large loops, while for odd occupation the loops are smaller and intersect inside the normal state CB diamonds. This dispersion is characteristic for ARs with a relatively small coupling to S. For example, comparing the extracted system parameters for resonance 2 results in $\Delta/\Gamma_{\rm S}\approx 5.5$, i.e. the large-$\Delta$ limit, in which $\Gamma_\mathrm{S}/U\approx 0.07 \ll 0.5$ suggests a gate tunable GS transition, consistent with the loop structure observed in the experiments. We note that also the strong-coupling expression predicts a GS transition, $k_{\rm B}T_{\rm KS}/\Delta\approx 1.7\times 10^{-3}\ll C$ for all $C$ in the literature. Similar results for the other resonances are summarized in Tab.~\ref{tab:ABS-devices} and are also consistent with the observed loop structures.

The full width at half the maximum (FWHM) of the three ARs in this device is $\sim 50-70\,\mathrm{\mu eV}$, much smaller than the total coupling strength $\Gamma \sim 0.2\,$meV (see Fig.~\ref{fig:ABS-A-Fits} in Appendix~\ref{app:SupplABS}), but consistent with the coupling $\Gamma_\mathrm{N} \approx 50\,\mathrm{\mu eV}$ to the two normal leads. This demonstrates that $\Gamma_{\rm N}=\Gamma_1+\Gamma_2$ fully accounts for the ABS excitation life time, as expected for $\Gamma_{N}\ll\Delta$, which holds for all devices investigated here.

\subsection{Device B}

\begin{figure}[t]
	\begin{center}
	\includegraphics[width=\columnwidth]{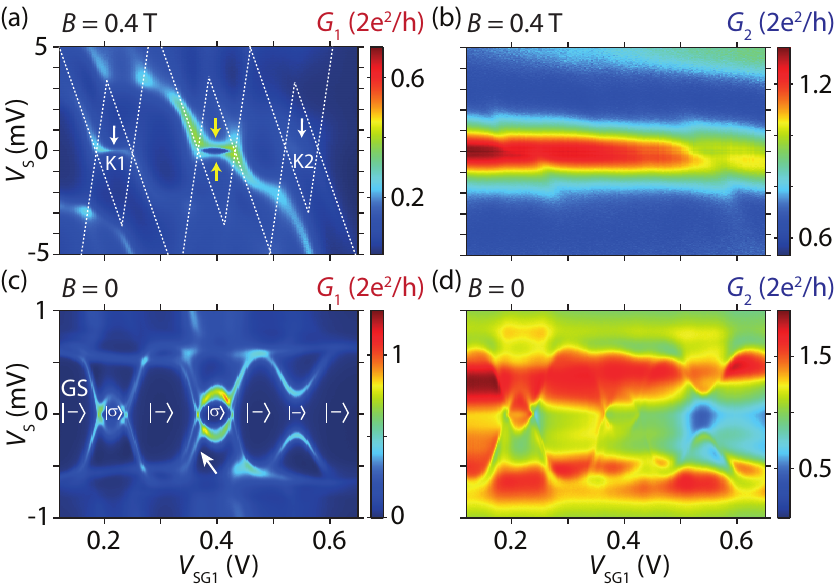}
	\end{center}	
	\caption{(Color online) $G_1$ and $G_2$ of device B as a function of $V_\mathrm{S}$ and $V_\mathrm{SG1}$ at $V_{21}=V_\mathrm{BG}=V_\mathrm{SG2}=0$, for $B=0.4\,$T (a-b) and $B=0$ (c-d). The white arrows in (a) label Kondo (K) ridges, and dashed lines qualitatively sketch the CB diamonds, where small (large) diamonds correspond to odd (even) occupation. GS denotes the QD groundstate in (c), and the white arrow points to a second Andreev resonance visible at a higher bias $|V_\mathrm{S}|$.}
	\label{fig:ABS-B-characterization}
\end{figure}

In device B a single QD is formed mainly between S and N1, while the CNT segment between S and N2 shows an essentially gate independent conductance in an open regime and can be viewed as a contact to the QD. This is shown in more detail in App.~\ref{app:SupplABS}. In Fig.~\ref{fig:ABS-B-characterization} we therefore plot $G_1$ and $G_2$ as a function of $V_\mathrm{S}$ and $V_\mathrm{SG1}$ to keep the other side unperturbed. In the normal state of S, $G_1$ in Fig.~\ref{fig:ABS-B-characterization}(a) exhibits a pattern typical for a QD strongly coupled to the leads, with an even-odd filling that can still be deduced from cotunneling lines. The CB diamonds are outlined by dashed lines. In contrast, $G_2$ plotted in Fig.~\ref{fig:ABS-B-characterization}(b) shows only a small conductance modulation, as expected for highly transparent barriers. In $G_1$ we find Kondo ridges in the odd charge states (K, arrows), which are strongly  broadened so that the Zeeman splitting cannot be resolved. The horizontal lines at $V_\mathrm{S}\sim \pm 0.2\,$mV in the middle of the conductance map (yellow arrows) do not agree with the expected Zeeman splitting of a (spin) Kondo resonance and $g=2$ and are probably due to inelastic cotunneling, or due to the valley Kondo effect.\cite{Herrero:2005} From the charge stability diagram in Fig.~\ref{fig:ABS-B-characterization}(a)  we estimate a leverarm of $\alpha_\mathrm{SG1}\approx 0.05$, a charging energy of $U\approx 3.0\,$meV, and a level spacing of $\sim 2.5\,$meV, consistent with a QD smaller than in device A and located between S and N1. Due to the pronounced Kondo resonances and the strong coupling to the leads, it is difficult to extract the tunnel coupling constants directly. We therefore estimate the coupling strength $\Gamma$ from the half-width at half maximum (HWHM) of the Kondo ridges K1 and K2.\cite{Buitelaar:2002,Deacon:2010,Kim:2013} We obtain $k_{\rm B}T_\mathrm{K1}\approx 0.24\,$meV ($T_\mathrm{K1}\approx 2.8\,$K) for K1, and estimate $\Gamma_\mathrm{K1} \approx 0.8\,$meV from $k_{\rm B}T_\mathrm{K}\approx\sqrt{\frac{U \Gamma}{2}}\exp\left(-\frac{\pi U}{8\Gamma} \right)$.\cite{Kim:2013} Similarly, for K2 we find $k_{\rm B}T_\mathrm{K2}\approx 0.8\,$ meV ($T_\mathrm{K2}\approx 9.3\,$K) and $\Gamma_\mathrm{K2} \approx 1.7\,$meV. These values are consistent with the CB resonance width at finite bias and show that the Zeeman splitting of the Kondo resonance can be neglected in this estimate.

$G_1$ and $G_2$ for S in the superconducting state are plotted in Figure~\ref{fig:ABS-B-characterization}(c-d), where we find pronounced ARs in $G_1$ for bias voltages $V_\mathrm{S}$ below the superconducting transport gap, $\Delta\approx 0.65\,$meV. A weak modulation at the same voltages can also be found in $G_2$. The AR loop in the left-most odd state (K1 in the normal state) is considerably smaller than the loops in device A, and the right-most AR loop (K2 in the normal state) does not exhibit a GS transition at all. The AR between K1 and K2 (arrow) exhibits two resonances related to QD excited states and consistent with normal state cotunneling lines, as discussed in more detail in Sec.~\ref{sec:ABS-Addbias-ExcitedAR}.

\begin{table*}
	\centering
	\begin{tabular}{|c||c|c|c|c|c|c|c|c|}
\hline 
\rule[-1ex]{0pt}{2.5ex} device & $\Delta$ (meV) & $U$ (meV) & $\Gamma_\mathrm{N}$ ($\mu$eV) & $\Gamma_\mathrm{S}$ ($\mu$eV) & $\Delta/\Gamma_{\rm S}$ & $\Gamma_\mathrm{S}/U$ & $k_{\rm B}T_{\rm KS}/\Delta$ & GS transition\\ 
\hline 
\rule[-1ex]{0pt}{2.5ex} A & 0.95 & 2.5 & 54/47/33 & 150/173/118 & 6.3/5.5/8.1 & 0.06/0.07/0.05 & $<2\times 10^{-3}$ & Y/Y/Y\\ 
\hline 
\rule[-1ex]{0pt}{2.5ex} B & 0.65 & 3.0 &  100/200 & 700/1500 & 0.93/0.43 & 0.23/0.50 & 0.29/1.05 & Y / N\\ 
\hline 
\rule[-1ex]{0pt}{2.5ex} C & 0.42  & 2.4 & $ 100/100$ & $700/780$ & 0.60/0.54 & 0.29/0.33 & 0.57/0.69 & crit. / N\\ 
\hline 
\end{tabular} 
	\caption{Extracted parameters and observed ABS loop structure for devices A, B and C. The values for different resonances are separated by a dash. Yes is abbreviated by Y, No by N. Crit. means that the AR loops touch.}
	\label{tab:ABS-devices}
\end{table*}

We now use the AR linewidth to estimate the remaining tunneling parameters. Equating the AR linewidths with $\Gamma_{\rm N}=\Gamma_1+\Gamma_2$ as established for device A, we find for resonance K1 $\Gamma^{\rm K1}_{\rm N}\approx 100\,\mathrm{\mu eV}$ and $\Gamma^{\rm K1}_{\rm S}=\Gamma_{\rm K1}-\Gamma^{\rm K1}_{\rm N}\approx 0.7\,$meV, with a corresponding Kondo temperature for the S contact of $k_{\rm B}T^{\rm K1}_{\rm KS}\approx 190\,\mu$eV.  With $\Delta/\Gamma^{\rm K1}_{\rm S}\approx 0.93$, this resonance is neither in the large-$\Delta$, nor in the strong-coupling limit. However, both expressions still correctly predict a GS transition: $\Gamma_{\rm S}/ U\approx 0.23 < 0.5$ in the large-$\Delta$ limit and $k_{\rm B}T_{\rm KS}^{K1}/\Delta\approx 0.29 < C$ for all mentioned $C$ in the strong coupling limit.
Similarly, we find for resonance K2 $\Gamma^{\rm K2}_{\rm N}\approx 200\,\mu$eV and $\Gamma_\mathrm{S}^\mathrm{K2}\sim 1.5\,$meV, with $k_{\rm B}T^{\rm K2}_{\rm KS}=680\,\mu$eV.
Here, $\Delta/\Gamma_{\rm S}^{\rm K2}\approx 0.4$, on a scale on which the use of expression \ref{Kondo_Temp} is still acceptable. This yields $k_{\rm B}T_{\rm KS}^{\rm K2}/\Delta\approx 1.1>C$ for all mentioned $C$, consistent with a suppression of the GS transition. We note that with $\Delta/\Gamma_{\rm S}^{\rm K2}\approx 0.5$ the large-$\Delta$ limit would predict touching AR loops, on the boundary of the phase transition, in contradiction to the experiment. Qualitatively, the larger coupling of the K2 state results in the suppression of the GS transition, whereas a transition is expected for K1.

\subsection{Device C}

\begin{figure}[b]
	\begin{center}
		\includegraphics[width=\columnwidth]{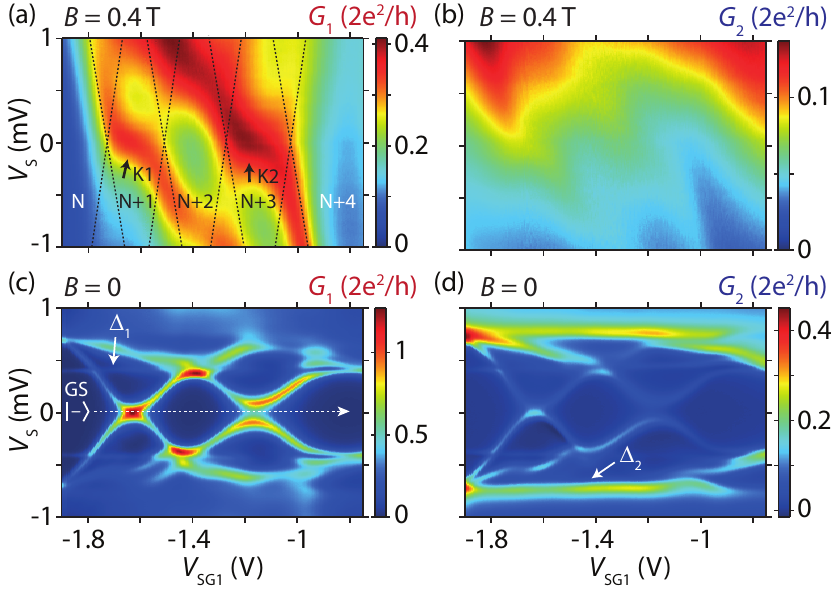}	
	\end{center}
	\caption{(Color online) $G_1$ and $G_2$ of device C as a function of $V_\mathrm{S}$ and $V_\mathrm{SG1}$ at $V_{21}=V_\mathrm{BG}=V_\mathrm{SG2}=0$, for $B=0.4\,$T (a-b) and $B=0$ (c-d). The black arrows in (a) label Kondo (K) ridges, and dashed lines sketch the CB diamonds. $\Delta_{1(2)}$ highlight the two superconducting transport gaps.}
	\label{fig:ABS-C-characterization}
\end{figure}

In contrast to the devices A and B, device C shows clear characteristics of a double QD with strong inter-dot coupling (see App.~\ref{app:SupplABS} for details). By only varying sidegate 1 and keeping $V_\mathrm{SG2}=0$, we mostly tune QD1 between N1 and S, while the conductance through QD2 between N2 and S is varied only around the maximum of a QD resonance (App.~\ref{app:SupplABS}). For this reason QD2 can essentially be seen as a contact to QD1. Figures~\ref{fig:ABS-C-characterization}(a-b) show $G_1$ and $G_2$ as a function of $V_\mathrm{S}$ and $V_\mathrm{SG1}$ for S in the normal state, with strong cotunneling lines  and Kondo resonances in the odd charge states (K1 and K2). From the charge stability diagram interpolated in Fig.~\ref{fig:ABS-C-characterization}(a) we estimate a leverarm of SG1 to QD1 of $\alpha_\mathrm{SG1}\approx 0.01$, a charging energy of $U\approx 2.4\,$meV and a level spacing of $\sim 1.3\,$meV for QD1. Similar as for device B, we estimate $\Gamma$ from the HWHM of the Kondo resonance and obtain $k_{\rm B}T_\mathrm{K1}\approx 0.3\,$meV ($T_\mathrm{K1}\approx 3.5\,$K) and $\Gamma_\mathrm{K1} \approx 0.8\,$meV for the resonance K1. For resonance K2 we find $k_{\rm B}T_\mathrm{K2}\approx 0.35\,$meV ($T_\mathrm{K2}\approx 4\,$K) and $\Gamma_\mathrm{K2}  \approx 0.88\,$meV.

For S in the superconducting state, $G_1$ and $G_2$ plotted in Figs.~\ref{fig:ABS-C-characterization}(c-d) both show pronounced ARs. In contrast to the previous devices, we find {\it two} horizontal conductance maxima, one at $\Delta_1\approx 0.42\,$meV most prominent in $G_1$, and $\Delta_2\approx 0.75\,$meV more clearly visible in $G_2$, which we tentatively interpret as two superconducting transport gaps in the individual CNT arms. Surprisingly, the ABS loops connect smoothly, but the right-most loops are bounded by $\Delta_1$, while the left-most clearly reaches up to $\Delta_2$. Currently we do not have a good explanation for these findings, but speculate that the QD wave function might be coupled inhomogeneously to S, e.g. at two separate places with two distinct effective coupling constants. For the present investigation the relevant findings are that the ABS loops exactly touch for the K1 state and that for K2 the loops are slightly separated in energy, i.e. that the K1 parameters correspond to the critical values of the ABS phase transition and that the GS transition is suppressed for K2.

The latter observations compare well to the extracted tunnel couplings:
For device C, we find an AR line width of $\Gamma_{\rm N}\approx 0.1\,$meV for both resonances and estimate $\Gamma_\mathrm{S}^{\rm K1}\approx 0.7\,$meV at the position of the Kondo resonance K1 and $\Gamma_\mathrm{S}^{\rm K2}\sim 0.78\,$meV for K2. Using $\Delta_1$ as the relevant energy gap, we find $\Delta_1/\Gamma_{\rm KS}^{\rm K1}\approx 0.60$ for K1 and $\Delta_1/\Gamma_{\rm KS}^{\rm K2}\approx 0.54$ for K2, both in a range for which it is acceptable to use Eqn.~(\ref{Kondo_Temp}). The latter results in $k_{\rm B}T_{\rm KS1}/\Delta_1\approx 0.57$ for K1, consistent with a `touching' of the AR loops if $C\approx 0.6$ in Eq.~(\ref{Kondo_Temp}), and $k_{\rm B}T_{\rm KS2}/\Delta_1\approx 0.69$ for K2, suggesting a close but avoided AR loop crossing and no GS transition. We note that the ratios used for the large-$\Delta$ limit, $\Gamma_\mathrm{S}^{\rm K1}/U\approx 0.29$ for K1 and $\Gamma_\mathrm{S}^{\rm K2}/U\approx 0.33$ for K2, would wrongly predict a GS transition for both charge states.

The results of this section are collected in Table~\ref{tab:ABS-devices} and compared to the AR loop structures found in the experiments. Our experiments clearly suggest a GS transition limit of $C\approx 0.6$ when using Eq.~(\ref{Kondo_Temp}). In addition, we find that the AR broadening is consistent with the coupling to both normal contacts.

\section{\label{sec:ABS-Addbias-ExcitedAR}Excited state Andreev resonances}

In Sec.~\ref{sec:ABS-devicecharacteriation} we found multiple replicas at higher energies for all ARs in device A and one replica in the central AR of device B. The spacings of $\sim 0.2-0.3$ and $\sim 0.5\,$meV between these replicas are very similar to the spacings of the lowest excited states in the normal state measurements of this device. We find no such replicas for device C, for which the orbital energies are larger than the superconducting gap, $\delta E > \Delta$.
We note that the replicas are not consistent with inelastic Andreev tunneling,\cite{GramichPRL:2015} because they show a clear gate dependence following the lowest AR, we find no negative differential conductance between the resonances, and they are broadened with increasing temperature. In the following, we exploit the third QD terminal to show that the replicas are caused by ABS mediated transport via excited states of the doublet $\ket{\sigma}$ in the odd QD state, a process we call excited state ARs.

\begin{figure}[b]
	\begin{center}
		\includegraphics[width=\columnwidth]{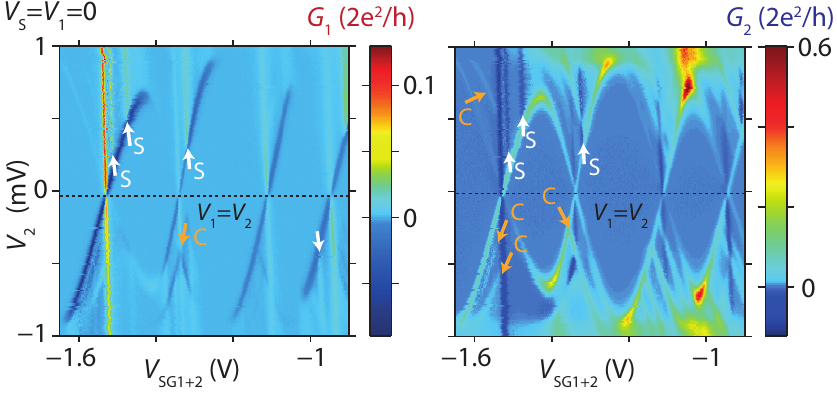}
	\end{center}
		\caption{(Color online) $G_1$ and $G_2$ of device A at $V_\mathrm{BG}=V_1=V_{\rm S}=0$ as a function of $V_2$ and a combined sidegate voltage $V_\mathrm{SG1+2}$ at $V_\mathrm{S, dc}=0$. The labels S and C denote partially overlapping straight Coulomb blockade resonances and curved (excited state) ARs, respectively. The horizontal dashed line marks the condition $V_2=V_1$. \cite{footnote1}}
	\label{fig:ABS-A-Exp-Offsetbias}
\end{figure}

To experimentally identify the origin of these replicas, we performed conductance measurements in a configuration different to the one in Sec.~\ref{sec:ABS-devicecharacteriation}. Here we keep the dc parts $V_1=V_{\rm S}=0$ and apply the dc bias $V_2$ to N2, while modulating $V_{\rm S}$ by the ac voltage $\delta V_{\rm AC}$ (see also Fig.~\ref{Fig1} for orientation). We again define the conductances $G_{1,2}=\delta I_{1,2}/dV_{\rm AC}$ from the current modulations in the normal terminals 1 and 2. Figure~\ref{fig:ABS-A-Exp-Offsetbias} shows $G_1$ and $G_2$ as a function of $V_\mathrm{SG1+2}$ and $V_{2}$. Intuitively, one might expect that $G_1$ is independent of $V_2$, which is the case for local processes when neglecting gating effects by $V_2$. Surprisingly, we find both, ABS mediated (`curved lines') and standard QD resonances (`straight lines') in these data sets. In the picture of $\Delta \rightarrow \infty$ this is not trivial and we will discuss this finding in more detail in Sec.~\ref{sec:ABS-FloatingS-Proximity}.

\begin{figure}[t]
	\begin{center}
	\includegraphics[width=\columnwidth]{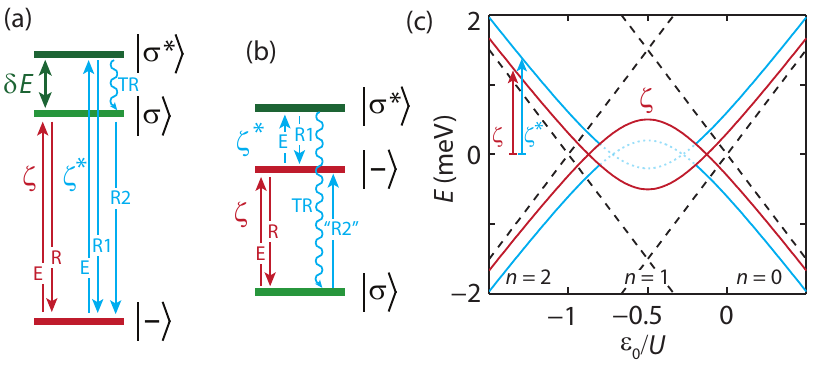}		
	\end{center}
	\caption{(Color online) (a-b) Expected low-energy charge transitions, including two spin-degenerate single-particle states $\ket{\sigma}$ and $\ket{\sigma^*}$ with a level spacing of $\delta E$ and the transition energies $\zeta$ and $\zeta^*$. TR abbreviates thermal relaxation. (a) shows the processes for $\ket{-}$ being the GS, (b) the ones for $\ket{-}$ set between $\ket{\sigma}$ and $\ket{\sigma^*}$. (c) $\zeta$ and $\zeta^*$ as a function of $\varepsilon_0/U$ for $U=3\,$meV, $\Gamma_\mathrm{S}=1\,$meV and $\delta E=0.3\,$meV. $n$ denotes the different charge states.}
	\label{fig:ABS-ExcitedAR}
\end{figure}

Here, we exploit that the width of both types of transport resonances between N1 and N2 is determined by $\Gamma_1+\Gamma_2$ only, and not by $\Gamma=\Gamma_1+\Gamma_2+\Gamma_\mathrm{S}$ as in the normal state (see Sec.~\ref{sec:ABS-devicecharacteriation}). This allows us to resolve also excited state resonances between the two N contacts. In Fig.~\ref{fig:ABS-A-Exp-Offsetbias} we find both, straight lines (labeled by white arrows) and AR with a curved dispersion (orange arrows), best visible in $G_2$. The straight excited state resonances have a similar spacing as the corresponding excited state ARs and sometimes even originate from the same position in these plots.

Excited state ARs can be understood on the same footing as the GS ARs in the $\Delta\rightarrow\infty$-limit (Sec.~\ref{sec:ABS_basics}). In addition to the odd-electron ground state $\ket{\sigma}$ we also consider the corresponding excited state $\ket{\sigma^*}$, separated in energy by $\delta E$. As depicted in Fig.~\ref{fig:ABS-ExcitedAR}(a) for $\ket{-}$ being the GS, the system can now also be excited to $\ket{\sigma^*}$ (blue arrows) in addition to $\ket{\sigma}$ (red arrows). The corresponding transition energies $\zeta^*$ (blue lines) and $\zeta$ (red lines) are plotted in Fig.~\ref{fig:ABS-ExcitedAR}(c). $\zeta^*$ replicates the shape of $\zeta$ at higher energies, as observed in the experiments. We call the resonances $\zeta^*$ \textit{excited state Andreev resonances}, because they correspond to a first order transition between an ABS and an excited odd-electron state. This situation of $\ket{-}$ being the GS results in the excited state AR highlighted by yellow arrows in the experiments of Fig.~\ref{fig:ABS-A-characterization}(c-d). Also consistent with the experiment is that the expected $\zeta^*$ line ends at the intersection with $\zeta$ in Fig.~\ref{fig:ABS-ExcitedAR}(b), best visible in the central resonance of Fig.~\ref{fig:ABS-B-characterization}(c). This can be understood qualitatively by introducing a (fast) energy relaxation from $\ket{\sigma^*}$ to $\ket{\sigma}$, which we call "thermal relaxation" (TR) to distinguish it from the parity changing relaxation by single electron tunneling. If $\ket{-}$ is the GS this results only in a second relaxation channel from $\ket{\sigma^*}$ to the GS, as illustrated in Fig.~\ref{fig:ABS-ExcitedAR}(a). However, if $\ket{\sigma}$ is the GS and $\ket{-}$ lies energetically midway between $\sigma^*$ and $\sigma$ or higher, as shown in Fig.~\ref{fig:ABS-ExcitedAR}(b), $\sigma^*$ can be excited by single electrons at $\zeta^*$, but the system relaxes fast to $\ket{\sigma}$, from where the energy $\zeta>\zeta^*$ is required to reach $\ket{-}$ to close the charge transport cycle. Therefore we expect for these gate voltages only one resonance at $eV_{\rm S}=\zeta$, but with an increased amplitude, because the additional transport channel via $\ket{\sigma^*}$ is allowed for $eV_{\rm S}\geq \zeta$.

\section{\label{sec:ABS-SerialTransport}ABS mediated transport between two N contacts}

In three-terminal QD devices, one can expect novel subgap transport mechanisms due to ``non-local'' transport processes that require two N and one S terminal, and their interplay with local effects involving only two terminals.\cite{Futterer:2009,Michalek:2015} In particular, transport between two N contacts can be mediated by ABSs.\cite{Michalek:2015} To perform this type of experiments, we use a different measurement configuration than in the previous sections, with the AC bias $\delta V_{\rm AC}$ and the dc bias $V_{1}$ applied to N1, while measuring the current variation $\delta I_2$ in N2. This can be done while also recording the current variation $I_{\rm S}$ in S at a fixed potential $V_{\rm S}$, or while leaving S floating (see Sec.~\ref{sec:ABS-FloatingS-Proximity}). We again define the differential conductances $G_{i}=\delta I_{i}/\delta V_{\rm AC}$.
		
\begin{figure}[b]
	\begin{center}
	\includegraphics[width=\columnwidth]{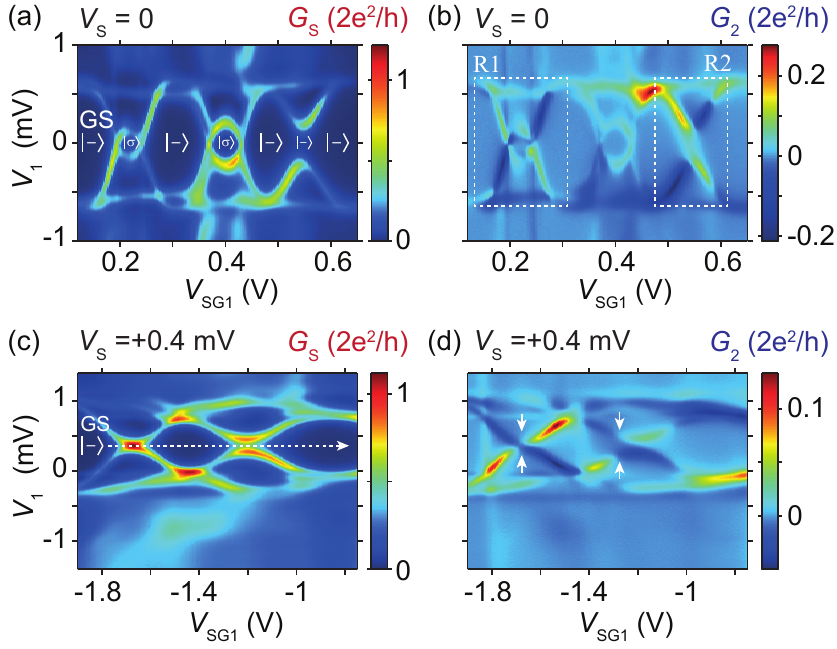}		
	\end{center}
	\caption{(Color online) $G_\mathrm{S}$ and $G_2$ as a function of the applied bias $V_1$ and the gate voltage $V_\mathrm{SG1}$, for (a-b) device B and $V_\mathrm{S}=0$, (c-d) device C and $V_\mathrm{S}=+0.4\,$mV. Arrows and dashed rectangles mark regions of pronounced conductance sign changes.}
	\label{fig:ABS-Serial-SgnChange}
\end{figure}
		
In Fig.~\ref{fig:ABS-Serial-SgnChange}(a-b) we plot $G_\mathrm{S}$ and $G_2$ of device B as a function of $V_{1}$ and the gate voltage $V_\mathrm{SG1}$, with $V_\mathrm{S}=0$. In addition to the typical AR pattern in $G_\mathrm{S}$ already discussed for Fig.~\ref{fig:ABS-B-characterization}, we find in $G_2$ pronounced resonances at the same gate and bias voltages. This illustrates again that N2 is a third terminal to the QD with ABSs. In contrast to the ARs in $G_{\rm S}$, the AR amplitudes in $G_2$ show pronounced sign changes, from positive to negative differential conductance with similar magnitudes $\sim \pm 0.1\times \rm 2e^2/h$ at positions symmetric around the electron-hole (e-h) symmetry points. The sign is inverted at the e-h symmetry point, at the singlet-doublet GS transitions, and for a reversed bias. This is best visible in the regions R1 and R2 marked by rectangles in Fig.~\ref{fig:ABS-Serial-SgnChange}(b). The central GS transition deviates from this description, most likely due to the overlapping second AR (see Sec.~\ref{sec:ABS-devicecharacteriation}), but also exhibits systematic sign changes. For the devices A and C we find similar characteristics as for device B. As an example, we plot $G_\mathrm{S}$ and $G_2$ of device C in Fig.~\ref{fig:ABS-Serial-SgnChange}(c-d), where we added a DC bias $V_\mathrm{S}=+0.4\,$mV to S. This bias results in an offset of the features in $G_\mathrm{S}$ and $G_2$, where we again find the characteristic sign changes in $G_2$, labeled by white arrows in Fig.~\ref{fig:ABS-Serial-SgnChange}(d), identical to region R2 in Fig.~\ref{fig:ABS-Serial-SgnChange}(b). From these and a series of other measurements with different $V_\mathrm{S}$ we conclude that the sign changes are robust against $V_\mathrm{S}$ and that all measurements can be understood considering potentials relative to the one on S.

\subsection{\label{subsec:ABS-RateEQmodel}Rate equation model}

\begin{figure}[t]
	\begin{center}
		\includegraphics[width=\columnwidth]{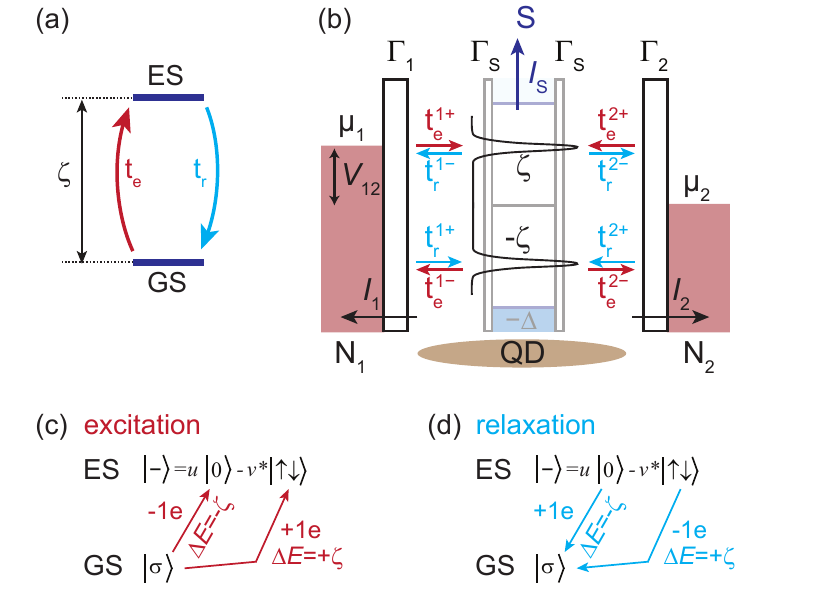}	
	\end{center}
	\caption{(Color online) (a) Two-level rate equation model with excitation (relaxation) rate $t_{\rm e}$ ($t_{\rm r}$). (b) Schematic of the three-terminal N-QD-S system, with a faint superimposed S contact, and a negative bias voltage $V_1$ applied to N1. Transport is only considered through the AR $\pm\zeta$, with color-coded rates $t_{\rm e(r)}^{1(2)+(-)}$. $I_1$, $I_2$ and $I_\mathrm{S}$ denote the technical current direction in the respective contacts (arrows).  (c-d)  Intuitive diagrams of the possible excitation (c, red) and relaxation (d, blue) processes and non-vanishing transition matrix elements, if $\ket{\sigma}$ is the GS. We consider only first order transitions (+1e/-1e) with the energy conditions $\Delta E = \pm \zeta$.}
	\label{fig:ABS-RateEQschem}
\end{figure}

To describe the transport through an S-QD system, we introduce a rate equation model similar to Ref.~\citenum{Schindele:2014}, using the $\Delta\rightarrow\infty$ limit to describe the ABSs and the doublet states, as shown in Fig.~\ref{fig:ABS-RateEQschem}(a). The excitation and relaxation rates depend on the transport processes, illustrated in Fig.~\ref{fig:ABS-RateEQschem}(b). In contrast to the two-terminal case, we now consider a coupling to a third normal metal terminal N2 with a strength $\Gamma_2$, in addition to the terminal S ($\Gamma_\mathrm{S}$) and N1 ($\Gamma_{1}$). For $E<\Delta$ we need to consider only transport mediated by the ABS. The bias $V_1$ is applied to N1 as in the experiment, while S and N2 are kept at $\mu_\mathrm{S}=\mu_2=0$. 

The excitation and relaxation rates, $t_{\rm e}$ and $t_{\rm r}$, denote the total rates by which the system is changed from the GS to the excited state (ES) and back, and will be determined below. The steady state occupation probabilities $P_{\rm GS}$ and $P_{\rm ES}=1-P_{\rm GS}$ of the GS and ES are then found by setting
\begin{equation}
\frac{d}{dt}P_{\rm ES}=t_{\rm e}P_{\rm GS}-t_{\rm r} P_{\rm ES}=0,
\label{eq:ABS-SteadyState}
\end{equation}
which yields $P_{\rm GS}=t_{\rm r}/(t_{\rm r}+t_{\rm e})$ and $P_{\rm ES}=t_{\rm e}/(t_{\rm r}+t_{\rm e})$. 

The rates $t_{\rm e}$ and $t_{\rm r}$ contain all excitation and relaxation processes, which can be viewed in the simplified picture of sequential individual tunneling events, as illustrated in Fig.~\ref{fig:ABS-RateEQschem}(b). In the corresponding rates we use the superscript $1 (2)$ to denote the contacts N1 (N2) and $+(-)$ to state whether an electron is added (removed) from the S-QD system. The excitation and relaxation rates can then be calculated using Fermi's Golden Rule.\cite{Schindele:2014,Braggio:2011,Wysokinski:2012} For example, if $\ket{\sigma}$ is the GS and $\ket{-}$ the ES, see Fig.~\ref{fig:ABS-RateEQschem}(c-d), the rate $t_{\rm e}^{1+}$ for the excitation by an electron tunneling in from N1 is 
\begin{equation}
\ket{\sigma}\xrightarrow[\rm N1]{+1e}\ket{-}:\; t_{\rm e}^{1+}=\Gamma_{\rm 1} 
\underbrace{\vert\braket{-|d_{\bar{\sigma}}^{\dagger}|\sigma}\vert^{2}}_{v^{2}} f_{1}(\zeta),
\end{equation}
where $f_{1 (2)}(E)$ is the Fermi distribution in contact N1 (N2), $d_{\sigma}^{\dagger}$ ($d_{\sigma}$) the creation (annihilation) operator for an electron on the QD with spin $\sigma$ (opposite spin $\bar{\sigma}$), and $v$ and $u$ are the BdG amplitudes of the ABS. Similarly, we find for the relaxation rate caused by electrons leaving to N2
\begin{equation}
\ket{-}\xrightarrow[\rm N2]{-1e}\ket{\sigma}:\; t_{\rm r}^{2-}=\Gamma_{\rm 2}
\underbrace{\vert\braket{\sigma|d_{\bar{\sigma}}|-}\vert^{2}}_{v^{2}}(1-f_{2}(\zeta)).
\end{equation}
We note that this latter process is possible only in a  three-terminal device. All such excitation and relaxation processes are depicted schematically in Fig.~\ref{fig:ABS-RateEQschem}(c-d), which allows to intuitively deduce all tunnel rates. The total excitation and relaxation rates then read
\begin{equation}
	\begin{split}
t_{\rm e}= &\underbrace{\Gamma_{1} v^2 f_{1}(\zeta)}_{t_{\rm e}^{1+}} + \underbrace{\Gamma_{2} v^2 f_{2}(\zeta)}_{t_{\rm e}^{2+}}\\
& + \underbrace{\Gamma_{1} u^2 (1-f_{1}(-\zeta))}_{t_{\rm e}^{1-}}+\underbrace{\Gamma_{2} u^2 (1-f_{2}(-\zeta))}_{t_{\rm e}^{2-}}
\\
t_{\rm r}= &\underbrace{\Gamma_{1} v^2 (1-f_{1}(\zeta))}_{t_{\rm r}^{1-}} + \underbrace{\Gamma_{2} v^2 (1-f_{2}(\zeta))}_{t_{\rm r}^{2-}}\\
& +\underbrace{\Gamma_{1} u^2 f_{1}(-\zeta)}_{t_{\rm r}^{1+}}+\underbrace{\Gamma_{2} u^2 f_{2}(-\zeta)}_{t_{\rm r}^{2+}}.
	\end{split}
\label{eq:ABS-rates}
\end{equation}
For the singlet $\ket{-}$ being the GS, all rates can be obtained from Eq.~(\ref{eq:ABS-rates}) by replacing $u$ by $v$, because the initial and final states in the transition matrix elements are interchanged. The total currents into N1 and N2 are then given by
\begin{equation}
	\begin{split}
&I_1 = \frac{e}{h}\left[ (t_{\rm e}^{1+} - t_{\rm e}^{1-}) P_{\rm GS} + (t_{\rm r}^{1+} - t_{\rm r}^{1-}) P_{\rm ES}\right]
\\
&I_2= \frac{e}{h}\left[ (t_{\rm e}^{2+} - t_{\rm e}^{2-}) P_{\rm GS} + (t_{\rm r}^{2+} - t_{\rm r}^{2-}) P_{\rm ES}\right],
	\end{split}
\label{eq:ABS-ModelCurrents}
\end{equation}
and $I_\mathrm{S}=-(I_1+I_2)$. The differential conductance measured at contact $i=\{1,2,S\}$ can then be obtained as $G_i = \mathrm{d}I_i/\mathrm{d}V_1$ for the experiments described in this section.

\begin{figure}[b]
	\begin{center}
	\includegraphics[width=\columnwidth]{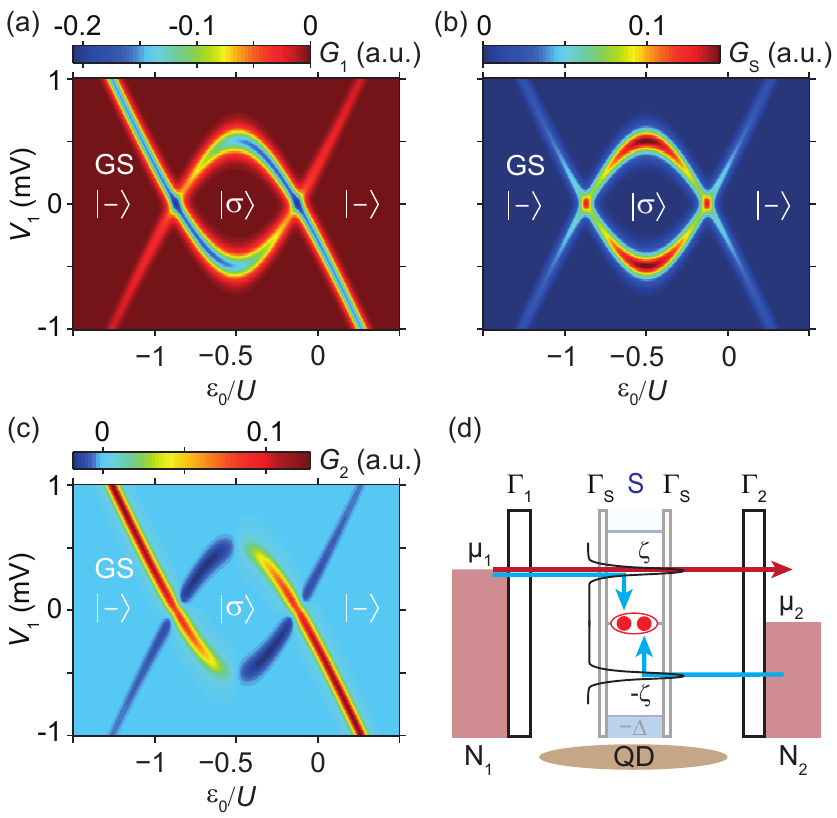}	
	\end{center}
	\caption{(Color online) (a-c) Model calculation of the differential conductance according to Eqs.~(\ref{eq:ABS-ModelCurrents}) and (\ref{eq:ABS-rates}) in the $\Delta\rightarrow \infty$ limit, for $\Gamma_\mathrm{S}=1\,$meV, $U=3\,$meV and $\Gamma_1=\Gamma_2=0.1\,$meV. (a) $G_1$, (b) $G_\mathrm{S}$ and (c) $G_2$ as a function of $V_1$ applied to N1 and the QD orbital energy $\varepsilon_0/U$, for $V_2=V_\mathrm{S}=0$, as in the experiment. To simulate a broadening of the AR, a small finite temperature of $T=0.5\,$K was assumed. (d) Schematic of the ``nonlocal'' transport processes for $\mu_1>+\zeta$. The red arrow depicts ``resonant ABS tunneling'', i.e. a direct transfer of one electron from N1 to N2 via the ABS, whereas the blue arrows show an inverse CPS process, i.e. the ``nonlocal'' creation of a Cooper pair in S.}
	\label{fig:ABS-RateEQmodel}
\end{figure}

\begin{figure*}[ht]
	\begin{center}
	\includegraphics[width=\textwidth]{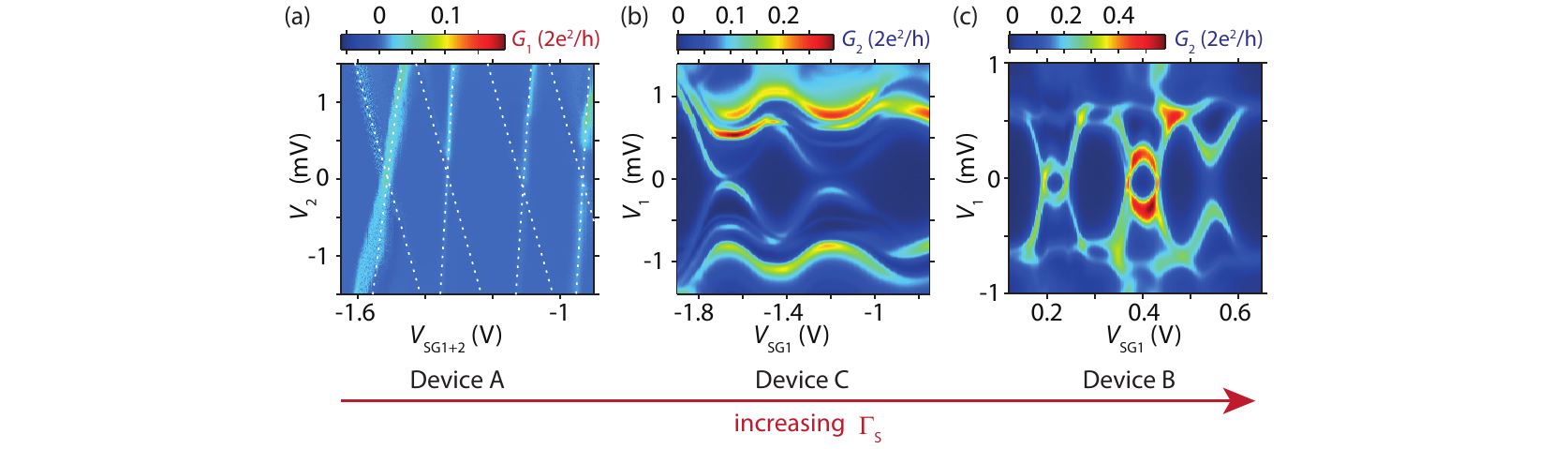}	
	\end{center}
	\caption{(Color online) Conductance between the two normal contacts N1 and N2 with the S contact floating. (a) $G_1(V_{21},V_\mathrm{SG1+2})$ for device A, (b) $G_2(V_{12},V_\mathrm{SG1})$ for device C, and (c) $G_2(V_{12},V_\mathrm{SG1})$ for device B. The measurements are ordered according to increasing coupling strengths $\Gamma_\mathrm{S}$.}
	\label{fig:ABS-Serial-floatingS}
\end{figure*}

An example of such a calculation is shown in Fig.~\ref{fig:ABS-RateEQmodel} for parameters that result in the familiar ABS loop structure and positive conductance values in the two-terminal local conductance $G_{\rm S}$. This is plotted in Fig.~\ref{fig:ABS-RateEQmodel}(b) as a function of $V_{1}$ and the normalized gate-tunable QD orbital energy $\varepsilon_0/U$ and corresponds, for example, to the resonances in the left-most region in Fig.~\ref{fig:ABS-Serial-SgnChange}(a). More importantly, Fig.~\ref{fig:ABS-RateEQmodel}(c) shows $G_2$ for the same parameter range, which should be compared to the left-most region R1 in Fig.~\ref{fig:ABS-Serial-SgnChange}(b). We find that the model at least qualitatively reproduces the experimental findings, with sign changes at the e-h symmetry point ($\varepsilon_0/U=-0.5$) and for a reversed bias, and clearly highlighting the GS transition. Similarly, for large $\Gamma_\mathrm{S}$ and $\ket{-}$ being the GS for all gate voltages, we find sign changes only at the e-h symmetry point, consistent with the resonances in the right-most region R2 in Figs.~\ref{fig:ABS-Serial-SgnChange}(b) and (d). The corresponding plots are shown in Appendix B in Fig.~\ref{fig:addedSim}. In addition, the sign changes in the model are independent of the applied voltages $V_\mathrm{S}$ or $V_2$, as in the experiment.

Due to the Fermi functions, the only non-vanishing rates at low temperatures are $t_{\rm e}^{1+}$, $t_{\rm r}^{1+}$, $t_{\rm r}^{2-}$ and $t_{\rm r}^{2+}$ for a forward bias $\mu_1>\mu_{\rm S}>\mu_2$, as depicted in Fig.\ref{fig:ABS-RateEQschem}. For example, the local Andreev processes between N1 and S scale with $t_{\rm e}^{1+}t_{\rm r}^{1+}$. More interestingly, the sign changes in $G_2$ are caused by the competition between processes comprising the two relaxation processes $t_{\rm r}^{2-}$ and $t_{\rm r}^{2+}$, which remove or add an electron from the system at different energies, respectively.
In this case the current in N2 is proportional to the difference between the BdG amplitudes, $I_2\propto (u^2-v^2)$, with a sign change at the GS transition (transition from $u^2>v^2$ to $u^2<v^2$), or for a reversed bias $V_1$. The conductance $G_2$ thus reflects directly the gradual charging of the ABS with decreasing gate voltage, which evolves from an excess charge of $0$ in the $n=0$ state to an average charge of $2e$ in the $n=2$ state. In a more coherent picture the process combining an electron tunneling into the system from both, N1 and N2, ($t_{\rm e}^{1+}$ combined with $t_{\rm r}^{2+}$) corresponds to the `nonlocal' creation of a Cooper pair in S, often called (inverse) Cooper pair splitting (CPS). Similarly, the coherent tunneling into and out of the system at the same energy ($t_{\rm e}^{1+}$ combined with $t_{\rm r}^{2-}$) corresponds to the direct transfer of one electron from N1 to N2, mediated by the ABS. This new process we call \textit{`resonant ABS tunneling'}. These `nonlocal' transport processes involve all three terminals and are illustrated in Fig.~\ref{fig:ABS-RateEQmodel}(d). For these processes we expect the rates $t_\mathrm{iCPS}\propto t_{\rm e}^{1+}t_{\rm r}^{2+}$ for inverse CPS and $t_\mathrm{rABSt}\propto t_{\rm e}^{1+}t_{\rm r}^{2-}$ for resonant ABS tunneling. This interpretation is also supported by recent more fundamental calculations,\cite{Futterer:2009,Michalek:2015} all in good agreement with our experiments and the simple model. We point out that to account for the similar magnitudes of the positive and negative differential conductance in $G_2$, it is necessary that CPS and resonant ABS tunneling are of similar strength.

\section{\label{sec:ABS-FloatingS-Proximity}Coherent oscillations in the ABS}

In this section we present and compare measurements with different boundary conditions imposed to the QD terminals. For example, Figs~\ref{fig:ABS-Serial-floatingS}(a-c) show the differential conductance $G$ between the two normal metal contacts (see also in the previous section) with the superconductor S left floating. The data for the devices A, C and B, are plotted as a function of the bias and gate voltage in the same gate voltage intervals as discussed in the previous sections. The data sets are ordered according to increasing $\Gamma_\mathrm{S}$, as found in Sec.~\ref{sec:ABS-devicecharacteriation}. Such experiments are equivalent to Ref.~\citenum{Higginbotham:2015}, where ABSs in a proximitized S-QD InAs nanowire were investigated in the context of Majorana bound states.

For device A, with a weak coupling to S and a small $\Gamma_\mathrm{S}/U$ ratio, we find CB diamonds very similar to the normal state. Specifically, there are no ARs, nor a suppression of $G$ related to a superconducting gap. ARs were, however, clearly observed when the superconductor was grounded (see Fig.~\ref{fig:ABS-A-characterization}), and both types of resonances were found in the experiments of Fig.~\ref{fig:ABS-A-Exp-Offsetbias}. In contrast, devices C and B, both with a larger coupling to S, show clear signatures of ARs and the energy gap. However, in these experiments no sign changes in $G$ occur and we observe only positive differential conductance features, in stark contrast to S being on a fixed potential (Fig.~\ref{fig:ABS-Serial-SgnChange}). For device B, this experiment results in essentially the same AR gate and bias dispersion as when measured between an N and the S contact (Fig.~\ref{fig:ABS-B-characterization}), whereas for device C we find additional curved resonances at higher energies (cf. Fig.~\ref{fig:ABS-C-characterization}). We note that the ARs are centered around zero bias, which suggests that $\mu_{\rm S}$ either follows $\mu_{\rm N1}$ or $\mu_{\rm N2}$ due to the different coupling strenghts to the normal metal terminals. 

The experiments on devices B and C are straight forward to understand qualitatively based on our previous picture of ABSs: we assumed that the CNT segment between N1 and N2 forms a QD in the normal state of S. In the superconducting state, the eigenstates of the segment are ABSs that determine all subgap transport. If S is at a fixed potential, it can take up (or emit) Cooper pairs, which allows Cooper pair splitting as a process and, in unison with resonant ABS tunneling, results in the pattern of positive and negative differential conductance. If S is floating, the experiments show that $\mu_{\rm S}$ is fixed, and since CPS and the local Andreev processes both add charges to S, which would change the electrical (though not the chemical) potential of S, these processes exactly compensate to satisfy the boundary condition $I_{\rm S}=0$. This leaves only resonant ABS tunneling as a subgap transport process between N1 and N2, which does not change the charge on S and results in a positive $G$, mapping out the ABS excitation energy. Additional features and different amplitudes compared to the traditional ARs are to be expected, since here only electrons at one energy are involved and we expect that the balance between the processes that change the charge on S depend on the bias.

The interpretation of the experiments on device A is less obvious. We find in the conductance between N1 and N2 the standard QD resonances when S is floating, while if S is at a fixed potential modulated by the ac voltage, ABS signatures occur simultaneously with the standard QD features, see Fig.~\ref{fig:ABS-A-Exp-Offsetbias} in Sec.~\ref{sec:ABS-Addbias-ExcitedAR}. In the two-terminal measurements between N and S, we only find ABS related signals. These experiments seem quite contradictory, since one would expect that the eigenstates of the S-QD system do not depend on the boundary conditions of the measurement. As discussed for samples B and C, the different processes are combined to satisfy the different boundary conditions, but the simultaneous observation of standard Coulomb blockade and ABS features cannot be resolved with such arguments alone.

This suggests that our picture of ABS mediated transport is not complete and only holds for a strong enough coupling to S. In this picture we assume that the QD develops new eigenstates due to the coupling to the unperturbed superconductor and one might expect that a continuous lowering of the coupling strengths results in a continuous transition from the ABS to the bare QD states. In contrast, our results seem to suggest that in some parameter regime both sets of states are available for different transport processes.

We speculate that for low enough $\Gamma_{\rm S}$ the tunneling of an electron on and off the QD from an N terminal is faster than the exchange of Cooper pairs with S, which is necessary for the hybridization of the QD with S. Right after the first tunneling event the system is in the state $\ket{\uparrow \downarrow}$, which is a coherent superposition of the system eigenstates $\ket{-}$ and $\ket{+}$, i.e., $\ket{\uparrow \downarrow}=-v\ket{-}+u\ket{+}$ ($\ket{0}=u^{*}\ket{-}+v^{*}\ket{+}$). Therefore we expect coherent oscillations between $\ket{0}$ and $\ket{\uparrow \downarrow}$ in an ABS excited by single electron tunneling events, with a characteristic frequency essentially determined by $\Gamma_{\rm S}$. This oscillation is expected to result in a suppression of the waiting time distribution for short times,\cite{Rajabi_Governale_PRL111_2013} and might be responsible for our findings here: in a three-terminal device an excitation cannot only be absorbed by forming a Cooper pair, as in a two-terminal device, but also by tunneling out into the third terminal (resonant ABS tunneling). If this happens faster than half a coherent oscillation, the tunneling electron stays in the bare QD state during the time it resides on the CNT segment. Assuming that an oscillation by half a period is required to transfer a Cooper pair to S, we find the relation $\Gamma_{\rm N2}>\Gamma_{\rm S}/\pi$ as a necessary condition to observe the bare QD resonances, in good agreement with device A, but not for the other two devices with considerably larger $\Gamma_{\rm S}$ and a faster oscillation rate relative to $\Gamma_{\rm N2}$. Our results demonstrate that a three-terminal QD can give access to the fast electron dynamics, in our example of the ABS coherent oscillations with frequencies $\Gamma_{\rm S}/h$ in the range between $\sim 30\,$GHz (device A) to $\sim 370\,$GHz (device B), by the implicit comparison to the tunnel rate to the third terminal.

\section{\label{sec:Conclusion}Conclusions}

In summary, we investigate electronic transport mediated by Andreev bound states in a three-terminal S-QD device and identify the coupling to the normal metal leads as the main source for the spectroscopic broadening of the Andreev resonances, and establish how the coupling to the superconductor determines the ground state of the system. In addition, we present `excited state Andreev resonances' at higher energies, with transitions between ABSs and odd parity excited QD states. We also report pronounced sign changes in the ABS-mediated transport between the two normal metal contacts, which we explain in an intuitive rate equation model as resulting from the competition between the nonlocal creation of a Cooper pair in S and a new process we call `resonant ABS tunneling', i.e., the subgap transport of single electrons through the S-QD system, only allowed in multi-terminal devices. Surprisingly, we find that depending on the imposed boundary conditions in the experiments it is possible to observe either ABSs or Coulomb blockade resonances, or both in the same experiment, which we tentatively attribute to the competition between coherent oscillations in the ABS (exchange of Cooper pairs with S) and the relaxation of the system by single electron tunneling into the second normal terminal.

We believe that experiments on complex quantum systems using multiple terminals provide a novel and clear experimental probe for many old and new phenomena, here for example the onset of the superconducting proximity effect on a QD and the formation of many-body quantum states. Such experiments also give direct access to the strength of the coupling between S and the QD, and a way to probe time dependent phenomena like the finite-frequency coherent oscillation between superposition states, without resorting to high-frequency and time domain experiments. In particular, we envisage experiments in similar structures with gate-tunable tunnel barriers\cite{Fulop_dHollosy_Baumgartner_PRB90_2014} to investigate in more depths the presented physical mechanisms.

\begin{acknowledgments}
We thank A. Levy Yeyati, J. Schindele and P. Makk for fruitful discussions. This work was financially supported by the Swiss National Science Foundation (SNF), the Swiss Nanoscience Institute (SNI), the Swiss NCCR QSIT, the ERC project QUEST and the EU FP7 project SE$^2$ND.
\end{acknowledgments}

\appendix

\section{\label{app:SupplABS}Additional data for device A, B and C}

In this appendix we analyze additional data of devices A, B and C. In particular, we discuss the electronic configuration of each device, using conductance maps as a function of both sidegate voltages. 

\subsection{Device A}

\begin{figure}[t]
	\centering
	\includegraphics[width=\columnwidth]{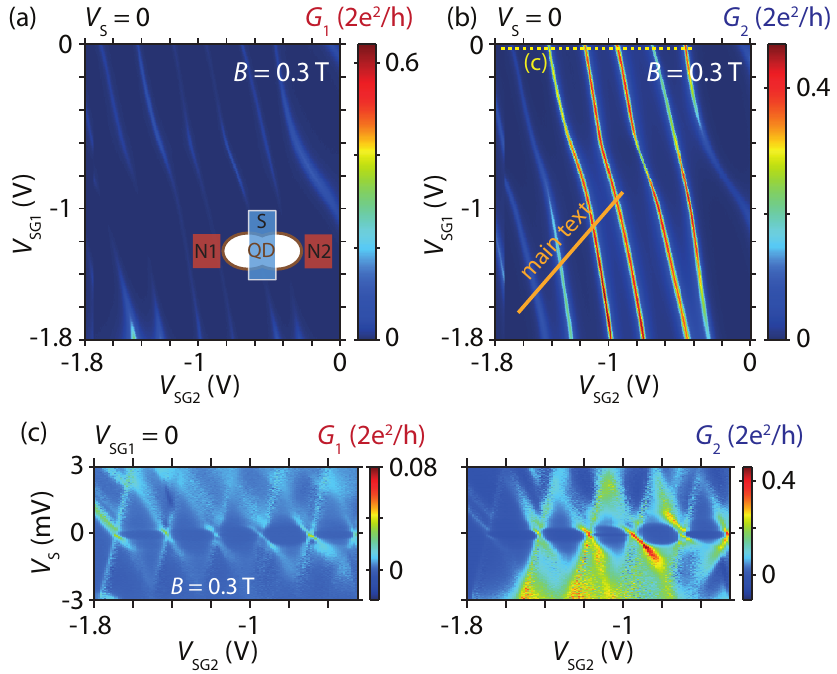}	
	\caption{(Color online) (a-b) Differential conductance $G_1$ and $G_2$ of device A as a function of the sidegate voltages $V_\mathrm{SG1}$ and $V_\mathrm{SG2}$, for $V_\mathrm{S}=V_\mathrm{21}=V_\mathrm{BG}=0$, at an external magnetic field of $B=0.3\,$T applied out-of-plane. The orange [dashed yellow] line in (b) indicate the studied gate voltages in the main text [in Fig.~(c)]. The inset in (a) sketches the assumed electronic QD configuration of the device. (c) $G_1$ and $G_2$ as function of $V_\mathrm{S}$ and $V_\mathrm{SG2}$ only, at $B=0.3\,$T and $V_\mathrm{SG1}=V_\mathrm{21}=V_\mathrm{BG}=0$.}
	\label{fig:ABS-A-OneQDobject}
\end{figure}

Figures \ref{fig:ABS-A-OneQDobject}(a) and (b) show the measured differential conductance $G_1$ and $G_2$ of device A as a function of the sidegate voltages $V_\mathrm{SG1}$ and $V_\mathrm{SG2}$, for the ac bias applied to S, but with $V_\mathrm{S}=V_\mathrm{BG}=0$, and in the normal state of the device at $B=0.3\,$T. These data are consistent with a single QD\cite{Livermore:1996,Wiel:2002, Fabian_Baumgartner_PRB_2016}, for example because we observe the same resonance lines in both, N1 and N2, and when we apply the ac bias to S. In addition, we find one dominant slope of the CB resonances that react similarly to both sidegates. There are only vague hints at a more complex confinement potential, possibly due to disorder or potential fluctuations on the substrate, e.g. a slight conductance modulation with some features stronger (weaker) in one (the other) arm, and slight changes in the slopes or the spacing between neighboring resonances. The electronic configuration we qualitatively deduce from these findings is sketched in the inset of Fig.~\ref{fig:ABS-A-OneQDobject}(a). In particular, in the region studied in the main text, marked by an orange line in Fig.~\ref{fig:ABS-A-OneQDobject}(b), the assumption of a single QD is justified. In Fig.~\ref{fig:ABS-A-OneQDobject}(c), we plot the differential conductance in the normal state of S along the yellow dashed line of Fig.~\ref{fig:ABS-A-OneQDobject}(b), as a function of the $V_\mathrm{S}$ and one sidegate voltage only. Here, we also clearly observe the same diamond structure in both arms of the device, further supporting this claim.

\begin{figure}[t]
	\centering
	\includegraphics[width=\columnwidth]{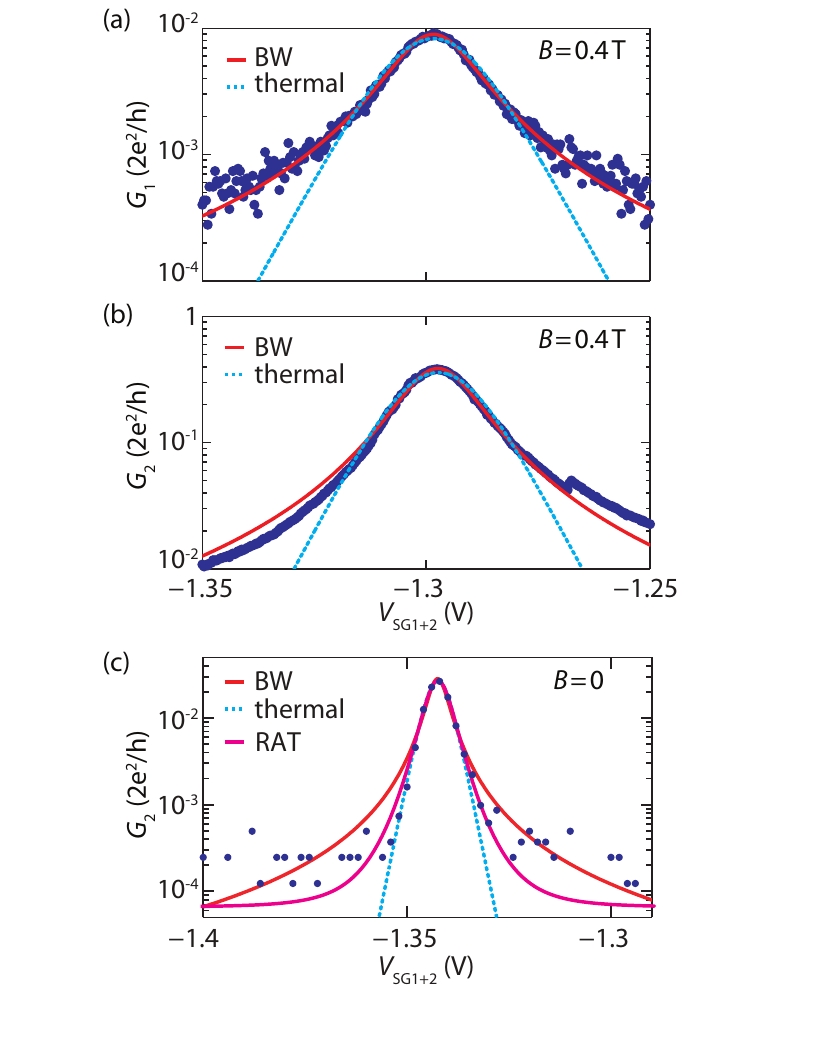}	
	\caption{Differential conductances $G_1$ (a) and $G_2$ (b) of resonance 1 (main text) as a function of the combined sidegate voltage $V_\mathrm{SG1+2}$, at $V_\mathrm{S}=V_\mathrm{21}=V_\mathrm{BG}=0$ and $B=0.4\,$T, i.e. when S is in the normal state. Red (blue dashed) lines represent the best fit obtained for a Breit-Wigner (BW) \cite{Leturcq:2004, Sanchez:2005} lineshape (thermally broadened) CB resonance. (c) $G_2$ of resonance 1 at $B=0$, i.e. when S is superconducting. In addition to fits with a BW and thermally broadened line we also show the fit to a two-terminal expression for resonant Andreev tunneling (RAT).\cite{GramichPRL:2015}}
	\label{fig:ABS-A-Fits}
\end{figure}

Next, we briefly demonstrate the fitting procedure to determine the individual $\Gamma_i$ of the contacts. In Fig.~\ref{fig:ABS-A-Fits}, the simultaneously measured zero-bias conductance $G_1\equiv G_\mathrm{1S}$ and $G_2\equiv G_\mathrm{2S}$ of resonance 1 (main text) is plotted as a function of the gate voltage $V_\mathrm{SG1+2}$ in the normal state of the device. Fits with a Breit-Wigner line-shape for a three-terminal device in the lifetime-broadened limit ($k_{\rm B}T\ll \Gamma$) and in the single-level transport regime ($k_{\rm B}T\ll U,\delta E$) with the gate-tunable position $E_0$ of the resonance, \cite{Leturcq:2004,Sanchez:2005} 
\begin{equation}
G_{ij}(E)=\frac{e^2}{h}\cdot \frac{\Gamma_i\Gamma_j}{(\Gamma_1+\Gamma_2+\Gamma_\mathrm{S})^2/4+(E-E_0)^2},
\end{equation}
agree very well with the data, and yield $\Gamma=\Gamma_1+\Gamma_2+\Gamma_\mathrm{S}\approx 205 \,\mathrm{\mu eV}$, $\Gamma_\mathrm{S}\Gamma_1\approx 198\,\mathrm{\mu eV}^2$ and $\Gamma_\mathrm{S}\Gamma_2\approx 7945\,\mathrm{\mu eV}^2$. From these equations, all parameters are determined for $\Gamma_\mathrm{S}> \Gamma_1,\Gamma_2$. This assumption can be directly justified from a similar analysis of measurements with the bias applied to N1, while measuring $G_\mathrm{S}\equiv G_\mathrm{S1}$ and $G_\mathrm{2}\equiv G_\mathrm{21}$ (not shown).
In addition, we show in Fig.~\ref{fig:ABS-A-Fits}(c) the zero-bias cross section of resonance 1 at $B=0$, i.e. for S being superconducting. As described in more detail before, \cite{GramichPRL:2015} the resonance is more narrow and better described by resonant Andreev tunneling (RAT). However, the expression used in Fig.~\ref{fig:ABS-A-Fits}(c) \cite{GramichPRL:2015} is meant for two-terminal devices, and accordingly the extracted parameters do not correspond in an obvious way to the ones in the normal state.

\subsection{Device B}

\begin{figure}[h!]
	\centering
	\includegraphics[width=\columnwidth]{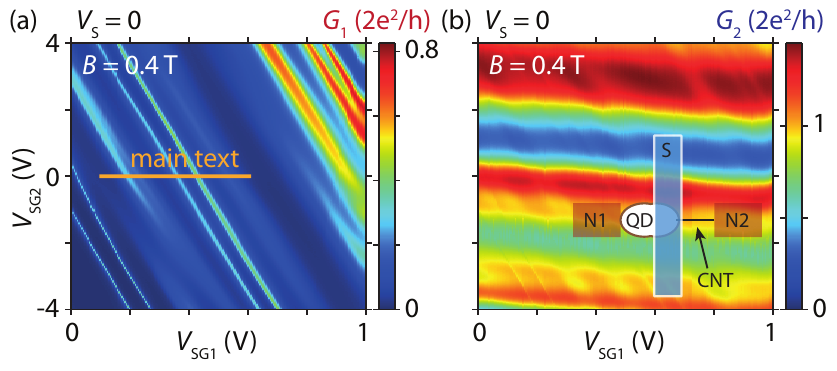}	
	\caption{(Color online) (a-b) $G_1$ and $G_2$ of device B as a function of the sidegate voltages $V_\mathrm{SG1}$ and $V_\mathrm{SG2}$, for $V_\mathrm{S}=V_\mathrm{21}=V_\mathrm{BG}=0$, at $B=0.4\,$T. The orange line in (a) indicates the studied gate voltage region in the main text. Inset in (b): Assumed electronic configuration of the CNT device.}
	\label{fig:ABS-B-GateGatemap}
\end{figure}

Figure \ref{fig:ABS-B-GateGatemap}(a) and (b) show $G_1$ and $G_2$ of device B as a function of the sidegate voltages $V_\mathrm{SG1}$ and $V_\mathrm{SG2}$, for the ac bias applied to S, but with $V_\mathrm{S}=V_\mathrm{BG}=0$, and in the normal state of the device at $B=0.4\,$T. In the conductance map of $G_1$, we observe a single dominant slope which is tuned mostly with $V_\mathrm{SG1}$, suggesting a QD located between S and N1. In contrast, $G_2$ plotted in Fig.~\ref{fig:ABS-B-GateGatemap}(b) shows a very high conductance with a weak and slow amplitude variation as function of $V_\mathrm{SG2}$, characteristic for a more open CNT regime with highly transmissive contacts. In particular, $G_2$ never approaches zero. Imprints of the resonances from $G_1$ can be observed in $G_2$, which we ascribe to resistive and capacitive cross-talk.\cite{Hofstetter:2009, Schindele:2012} Hence, we assume an electronic configuration of the CNT device as schemtically depicted in the inset of Fig.~\ref{fig:ABS-B-GateGatemap}(b), with a larger QD on the left side of the device mostly tunable by $V_\mathrm{SG1}$, and an `open' CNT lead to the right. 

\subsection{Device C}

\begin{figure}[b]
	\centering
	\includegraphics[width=\columnwidth]{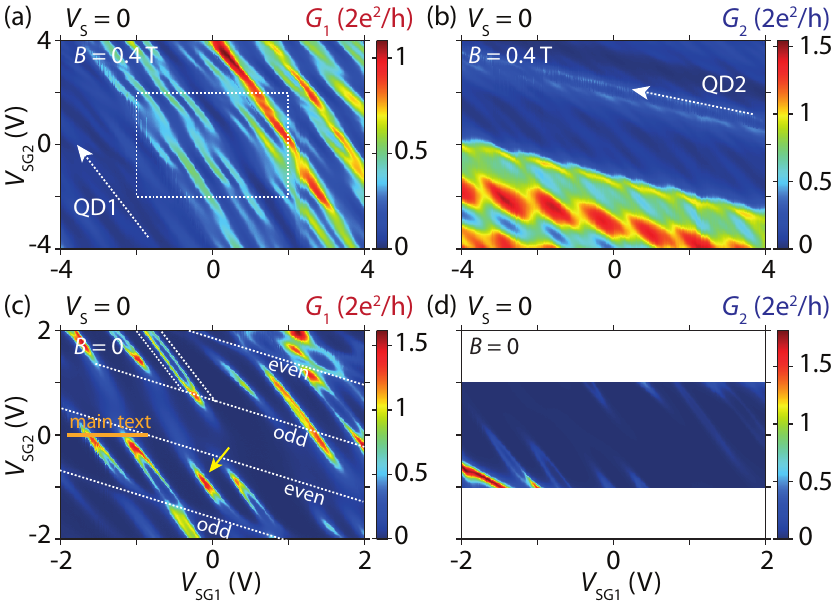}	
	\caption{(Color online) $G_1$ and $G_2$ of device C as a function of the sidegate voltages $V_\mathrm{SG1}$ and $V_\mathrm{SG2}$, for $V_\mathrm{S}=V_\mathrm{21}=V_\mathrm{BG}=0$, at $B=0.4\,$T (a-b) and at $B=0$ (c-d). White dashed arrows in (a,b) denote different slopes consistent with resonances of QD1 or QD2. The white rectangle marks the region studied in (c,d). In (c), the orange line corresponds to the gate voltage region studied in the main text, and dashed lines denote the charge states of QD2 (even/odd).}
	\label{fig:ABS-C-GateGatemap}
\end{figure}

In Fig.~\ref{fig:ABS-C-GateGatemap}, $G_1$ and $G_2$ of device C are plotted as a function of the sidegate voltages $V_\mathrm{SG1}$ and $V_\mathrm{SG2}$, for the ac bias applied to S, but with $V_\mathrm{S}=V_\mathrm{BG}=0$, for S in the normal state at $B=0.4\,$T (a-b) and in the superconducting state at $B=0$ (c-d).

In contrast to device A and B, a charge stability diagram with anti-crossings characteristic for a double quantum dot with strong inter-dot coupling and hybridization is observed in the normal state of S. In Fig.~\ref{fig:ABS-C-GateGatemap}(a) and (b), the dominant conductance lines with different slopes in $G_1$ and $G_2$ are consistent with resonances of QD1 or QD2. An apparent smearing of the resonances in every other charge state, both for QD1 and QD2, is due to pronounced Kondo ridges. In particular, the broad resonances in Fig.~\ref{fig:ABS-C-GateGatemap}(a) which we ascribe to QD1 indicate an odd charge state with a Kondo ridge. This is more obvious when one closely inspects the region indicated by a dashed rectangle in Fig.~\ref{fig:ABS-C-GateGatemap}(a) in the superconducting state of S, which is plotted in Fig.~\ref{fig:ABS-C-GateGatemap}(c). Here, one observes a pair of Andreev resonance lines in the even charge state of QD2 (yellow arrow), corresponding to the two singlet-doublet GS transitions observed in the odd charge states of QD1. These GS transitions vanish close to the boundary and in the odd charge state of QD2, which can for example be seen by following the Andreev resonances marked with a yellow arrow. In the main text, we focus on Andreev states in QD1 for a fixed even charge state of QD2, but close to its charge degeneracy point. The corresponding gate voltage region is indicated by an orange line in Fig.~\ref{fig:ABS-C-GateGatemap}(c).

\section{\label{sec:addedSim}Model results for large $\Gamma_{\rm S}$}

\begin{figure}[h]
	\begin{center}
	\includegraphics[width=\columnwidth]{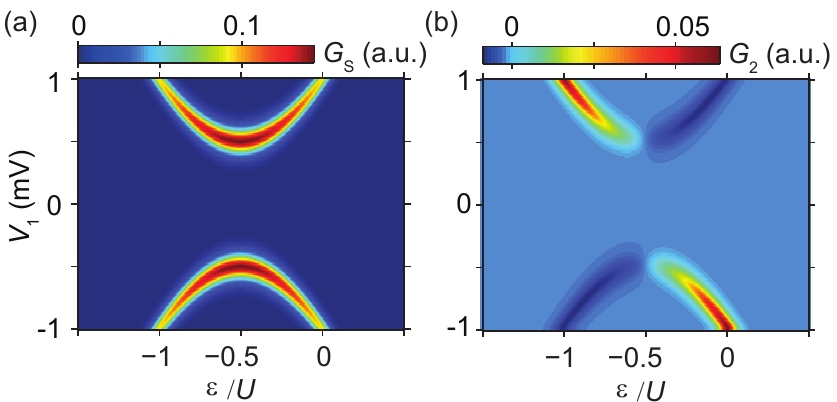}	
	\end{center}
	\caption{(Color online) Model calculation for $\Gamma_\mathrm{S}=2\,$meV, $U=3\,$meV and $\Gamma_1=\Gamma_2=0.1\,$meV. (a) shows $G_\mathrm{S}$ and (b) $G_2$ as a function of $V_1$ applied to N1 and the QD orbital energy $\varepsilon_0/U$, for $V_2=V_\mathrm{S}=0$, as in the experiment. To simulate a broadening of the AR, a temperature of $T=0.5\,$K was assumed.}
	\label{fig:addedSim}
\end{figure}

Figure~\ref{fig:addedSim} shows the model calculation of the differential conductance according to Eqs.~(\ref{eq:ABS-rates}) and (\ref{eq:ABS-ModelCurrents}) in the $\Delta\rightarrow \infty$ limit for the same parameters as in Fig.~\ref{fig:ABS-RateEQmodel}, but with a twice as large $\Gamma_{\rm S}$. This calculation reproduces at least qualitatively the measurements in region R2 of Fig.~\ref{fig:ABS-Serial-SgnChange}(b), which shows the data for a large-$\Gamma_{\rm S}$ AR.

\bibliography{Ref_Pb-3T-ABS} 

\end{document}